\documentclass[preprint,aps,nofootinbib]{revtex4}
\usepackage{dcolumn}
\usepackage{bm}
\usepackage{tabularx}
\usepackage{amsmath}
\usepackage{amscd}
\usepackage{delarray}
\usepackage{amssymb}
\usepackage[dvips]{graphicx}

\newcommand{\psla}{p\kern-.45em/}
\newcommand{\ebsla}{\bar{\epsilon}\kern-.45em/}
\newcommand{\qsla}{q\kern-.45em/}
\newcommand{\pbsla}{\bar{p}\kern-.45em/}
\newcommand{\qbsla}{\bar{q}\kern-.45em/}

\newcommand{\slush}{\!\!\!/}

\def\ti    {\tilde}

\def\sq    {{\ti q}}

\def\stau  {{\ti\tau}}

\def\ch    {\ti {\chi}}

\def\cpl    {\ti {\chi}^+}
\def\cm    {\ti {\chi}^-}

\def\nt    {\ti {\chi}^0}
\def\none  {\ti \chi^0_1}
\def\ntwo  {\ti \chi^0_2}
\def\nthre {\ti \chi^0_3}
\def\nfour {\ti \chi^0_4}

\def\sq    {\ti q}

\def\gluino{\ti g}

\def\h     {\frac{1}{2}}
\def\gsim  {\hspace{0.3em}\raisebox{0.4ex}{$>$}\hspace{-0.75em}\raisebox{-.7ex}{$\sim$}\hspace{0.3em}}
\def\lsim  {\hspace{0.3em}\raisebox{0.4ex}{$<$}\hspace{-0.75em}\raisebox{-.7ex}{$\sim$}\hspace{0.3em}}
\def\propsim  {\hspace{0.3em}\raisebox{0.4ex}{$\propto$}\hspace{-0.75em}\raisebox{-.7ex}{$\sim$}\hspace{0.3em}}

\begin{document}

\preprint{KEK-TH-1134}
\preprint{YITP-07-03}

\title{The study of $\tilde{q}_L \tilde{q}_L$ production at LHC in the $l^\pm l^\pm$ channel
\\
and sensitivity to other models}

\author{Mihoko M.Nojiri}
 \email{nojiri@post.kek.jp}
\author{Michihisa Takeuchi}%
 \email{tmichihi@post.kek.jp} 
\affiliation{$^{\ast\dagger}\!\!\!$ 
Theory Group, KEK,
and $^{\ast}\!\!\!$ the Graduate University for Advanced Studies (SOKENDAI)\\
1-1 Oho, Tsukuba, 305-0801, Japan\\
$^\dagger\!\!\!$ Yukawa Institute for Theoretical Physics, Kyoto University,\\
 Kyoto 606-8502, Japan
}

\date{\today}

\begin{abstract}
At the LHC,
$\tilde{q}_L \tilde{q}_L$ production is one of the main SUSY production processes
in the MSSM, which occurs due to the chirality flip caused by
the Majorana gluino mass. This process is one of the sources of
same sign two lepton (SS2$l$) events, however,  gluino production also
contributes to this channel.  In this paper,
we develop a method to identify gluino and squark production separately
in the SS2$l$ channel,  based on cuts
on the kinematical configuration of the jets.
We applied the method to the MSSM, a model with an extended gluino  
sector, and
the Littlest Higgs model with T-parity (LHT), and found a distinctive
difference between these models when considering the numbers of  
SS2$l$ events selected by the cuts.
\end{abstract}

\maketitle


\section{Introduction}
\label{sec:introduction}

The Standard Model (SM) describes interactions among elementary particles
very well.
However astrophysical and cosmological observations 
such as WMAP have confirmed the existence of 
dark matter (DM) that cannot be explained in the SM \cite{wmap}.
The particle which DM consists of does not leave detectable signals in the detectors
at high energy collider experiments,
because it should be weakly interacting.

If new particles are produced at collider experiments 
and decay into visible particles 
and a DM particle,
we can observe large missing transverse momentum ($E\slush_T$) in the events.
Many models which predict DM candidates have been proposed.
Among them,
the minimal supersymmetric standard model with conserved R-parity (MSSM)
is an attractive one.
The lightest supersymmetric particle (LSP) is stable and a DM candidate.
In the MSSM, quadratic divergences 
in the Higgs mass radiative corrections cancel each other,
therefore the fine tuning problem is solved.
At the large hadron collider (LHC), a $pp$ collider with $\sqrt{s}=14\,$TeV
starting its operation in 2007 \cite{atlas,cms},
discovery of the squarks $\sq$ and gluino $\gluino$
-- super partners of quarks and gluon --
is possible for the masses up to 2.5\,TeV by looking for an excess of events 
with large $E\slush_T$. 
We can also measure their mass spectrum
by studying the decay kinematics of the quarks and gluino decay chains
if there are enough events.

However,
the discovery of a $E\slush_{T}$ signature does not necessarily mean
the confirmation of the existence of supersymmetry.
A similar mass spectrum and decay pattern 
might be obtained 
for the Universal Extra Dimension model (UED) \cite{UED}
and 
the Littlest Higgs model with T parity (LHT) \cite{Cheng:2003ju,Cheng:2004yc,Hubisz:2004ft}.
To study the origin of the $E\slush_{T}$ signature,
therefore,
it is important 
to measure the other features 
that are characteristic of the MSSM. 

Many analyses have already been carried out in this direction.
Recently, 
processes sensitive to the $\gluino\sq q$ Yukawa type coupling constant 
 have been investigated in \cite{Freitas:2006wd}.
The Yukawa type coupling constant for the $\gluino\sq q$ vertex
is the same as the gauge coupling constant due to SUSY.
They study the same sign two isolated-lepton (SS2$l$) channel
to estimate the production cross section $\sigma(\sq_L\sq_L)$. 
The SS2$l$ channel is one of the major discovery channels for supersymmetry,
which is studied in \cite{Baer:1990fb,Baer:1991xs,Baer:1995va}.
The process is in principle sensitive to the coupling,
however
they found a large background from 
$\gluino \sq$ production. 
To measure the coupling constant,
it is important to measure the cross sections of sparticle
production processes separately. 
In this paper, we also focus on 
the $\sigma(\sq_L\sq_L)$.
We will give a new method based on cuts on the numbers of jets 
in the hemispheres for the purpose of separating $\sq_L\sq_L$ production 
from $\gluino \sq_L$ production. 

The $\sq_L\sq_L$ production process
occurs through a chirality 
flip caused by the gluino majorana mass term $m_g \gluino\gluino$.
To study the sensitivity to the majorana nature of the gluino mass,
we consider a model with an extended gluino sector.
This extension is inspired by the model that extends SUSY to $N=2$ 
in \cite{Fox:2002bu,Chacko:2004mi,Carpenter:2005tz,Hisano:2006mv}.
In this model,
an adjoint matter $\ti a$ is introduced,
then the gluino can have a Dirac mass term $m_D \gluino \ti a$.
The gluino mass receives a contribution from $m_D$.
We also discuss the Littlest Higgs model with T-parity (LHT).
This model contains quark partners ($q_-$) and gauge boson partners ($W_H$, $A_H$) 
which decay into the stable lightest T-odd particle (LTP)
$A_H$ and SM particles as in the MSSM.
Then, collider signatures are similar.
There is progress concerning spin studies at LHC
to distinguish these models \cite{Barr:2004ze,Athanasiou:2006ef}.
In this paper, however,
we focus only on the difference among 
the production cross sections in these models.

This paper is organized as follows.
In Section 2, we first discuss the mass dependencies of 
the production cross section of $\sq$ and $\gluino$ in the MSSM.
The $\ti u_L\ti u_L$ production cross section is 
typically of the order of $100\,$fb at $m_{\sq} \sim m_{\gluino} \sim 1\,$TeV,
which may be detectable at LHC.
The production processes of $\sq_L\sq_L$ and $\sq_R\sq_R$ 
occur due to the gluino majorana mass,
therefore the mass dependencies are different from 
those of $\sigma(\sq_L\sq_R)$ and $\sigma(\sq_L\sq_L^\ast)$. 
We compare $\sigma(\sq_L\sq_L)$ with $\sigma(\gluino\sq_L)$
and find that $\sigma(\sq_L\sq_L) \ll \sigma(\gluino\sq_L)$.
In this section,
we also choose a few model points for later analyses.

For the MSSM with an extended gluino sector,
the $\gluino$, $\sq$ production cross sections are functions of 
the three mass parameters
-- two majorana masses $m_g,m_A$ and one Dirac mass $m_D$,
where the mass terms are of the form $m_g \gluino \gluino + m_A \ti a\ti a 
+ m_D(\ti a \gluino + \gluino \ti a$).
In particular, $\sigma(\sq_L\sq_L)$
 becomes zero for some particular choice of parameters.
We also discuss the LHT model. The production process 
$q_- q_-$ occurs by heavy SU(2) gauge boson exchange.
$\sigma(q_- q_-)$ is large compared with $\sigma(\sq\sq)$ of the MSSM \cite{Belyaev:2006jh}.
On the other hand,
because no gluon partner exists in this model,
there is no problematic background corresponding to the $\gluino\sq$ production of the MSSM. 

In Section 3,
We study SS2$l$ events
to estimate $\sigma(\sq_L\sq_L)$ in the MSSM.
To reduce SS2$l$ events from $\gluino\sq_L$,
we use a $b$-jet veto and the numbers of jets in hemispheres 
defined following the procedure proposed in \cite{hemisphere}.
We demonstrate that production processes
$\gluino\gluino,\gluino\sq,\sq\sq$ can be distinguished 
by the efficiency 
under the cuts.
We also discuss the dominant $t\bar{t}$ background.

In Section 4,
we calculate the expected number of SS2$l$ events 
in the model with an extended gluino sector considered in Section 2.
We discuss the sensitivity of the $\sq_L\sq_L$ production cross section
to the majorana gluino mass $m_g$
at a few model points. 
We also estimate the number of SS2$l$ signature in the LHT model.
The efficiency under the cuts to reduce gluino background turns out 
to be useful to prove the existence of quark partner productions
and difference from the MSSM prediction. 
Section 5 is devoted to the summary.

\section{Production cross sections at LHC}
\subsection{MSSM production cross sections at LHC}
\label{MSSMprod}
In the MSSM,
sparticles are always pair produced at LHC because of R-parity conservation.
Production processes $\gluino \gluino$, $\gluino \sq$, $\sq\sq$
occur copiously unless the masses are much heavier than 1\,TeV.
Because $u$ quark and $d$ quark parton distribution functions (PDF) of a proton
are much harder than the other partons,
$\ti u$ and $\ti d$ are mainly produced among squarks.
In particular, production processes 
$\ti u_L \ti u_L$, $\ti d_L \ti d_L$, 
$\ti u_R \ti u_R$, $\ti d_R\ti d_R $, etc.
require a chirality flip, 
therefore they do not occur if the gluino mass is not of majorana type $m_g \lambda\lambda$.
Only gluino exchange diagrams contribute to the productions.\footnote{
Here, we neglect the contributions 
from neutralino and chargino exchange diagrams.
This assumption is reasonable, because 
$g^2<g_s^2, m_{\ti W} < m_{\gluino}$ in mSUGRA.}

\begin{figure}[htbp]
\begin{minipage}{8cm}
\includegraphics[scale=0.9]{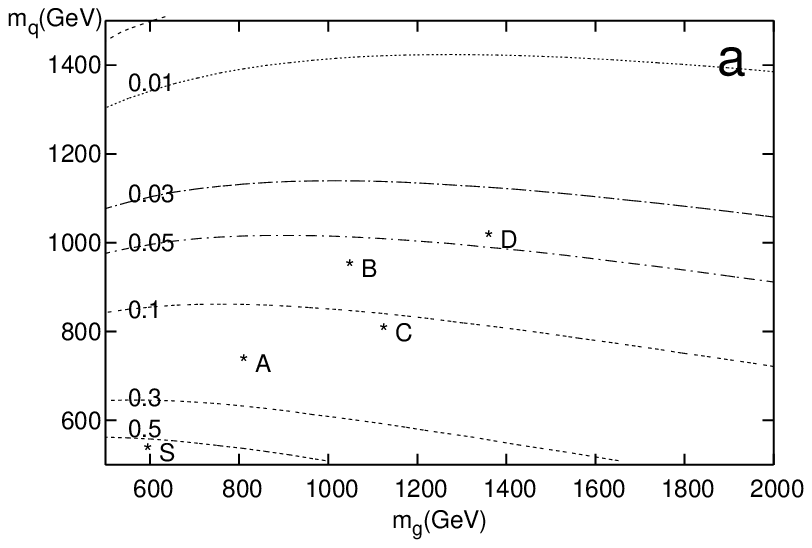}
\end{minipage}
\hfill
\begin{minipage}{8cm}
\includegraphics[scale=0.9]{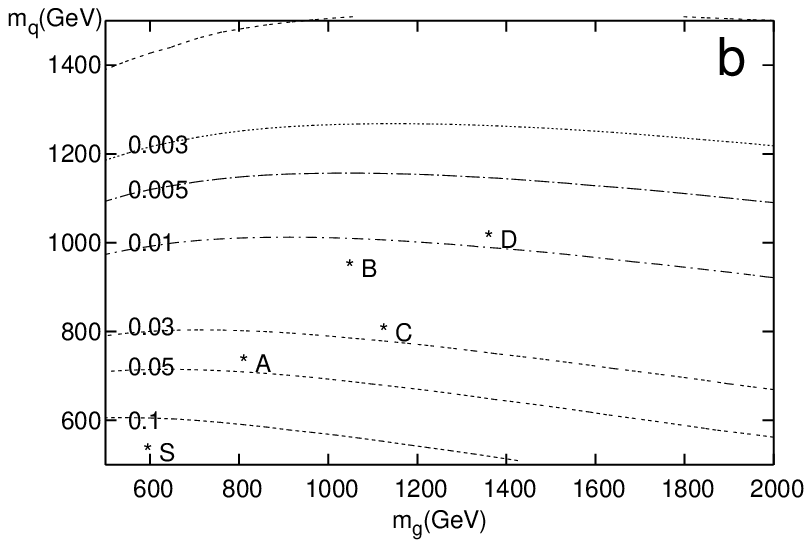}
\end{minipage}
\\
\begin{minipage}{8cm}
\includegraphics[scale=0.9]{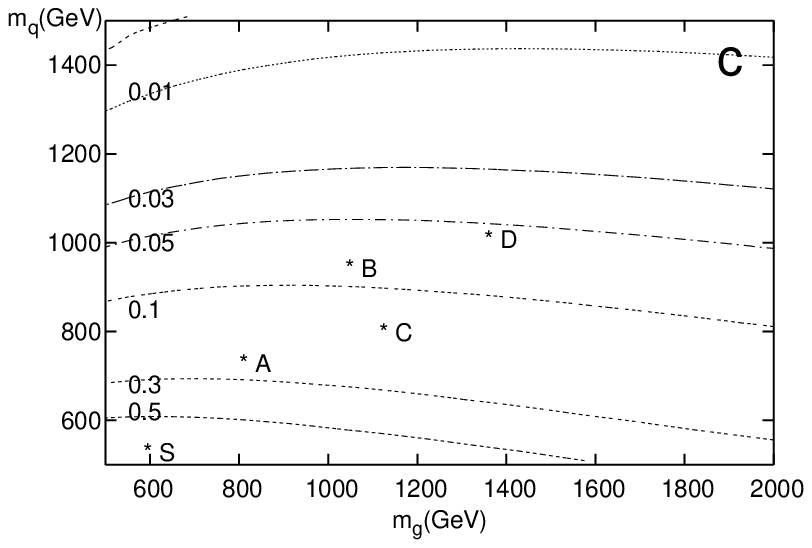}
\end{minipage}
\hfill
\begin{minipage}{8cm}
\includegraphics[scale=0.9]{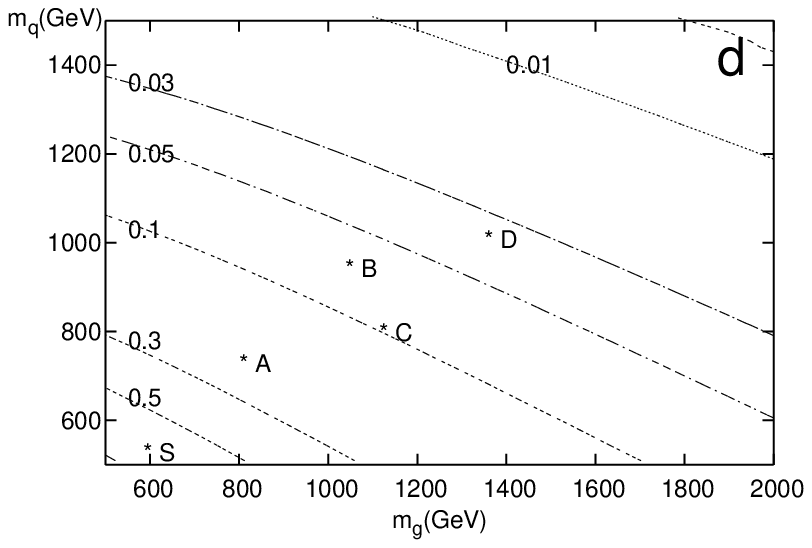}
\end{minipage}
\caption{Contour plot of a) $\sigma(\ti u _L \ti u _L)$,
b) $\sigma(\ti d _L \ti d _L)$,
c) $\sigma(\ti u _L \ti d _L)$ and
d) $\sigma(\ti u _L \ti u _R)$ as a function of 
$m_{\gluino}$ and $m_{\sq}$.}
\label{f1}
\end{figure}

\begin{figure}[htbp]
\center{
\begin{minipage}{8cm}
\includegraphics[scale=0.9]{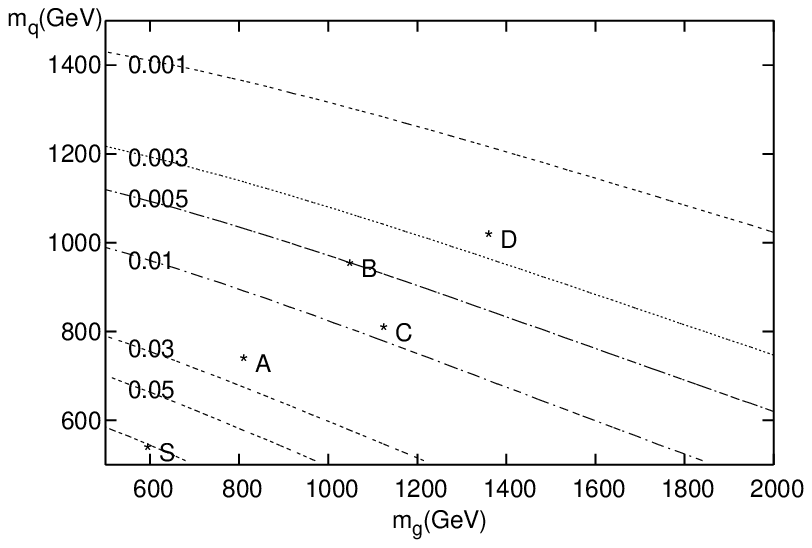}
\end{minipage}
\caption{Contour plot of $\sigma(\ti u_L \ti d _L^\ast)$ as a function of 
$m_{\gluino}$ and $m_{\sq}$.}
\label{f3}
}
\end{figure}

If the sparticle mass spectrum is known, 
the production cross sections at LHC are calculable. 
The mass dependencies of some sparticle production cross sections
are shown  
in Figures \ref{f1}$\sim$\ref{f3}.
Here, we use CTEQ 6l~\cite{CTEQ6l} as the PDF.
The horizontal axis is the gluino mass $m_{\gluino}$,
and the vertical axis is the squark mass $m_{\sq}$.\footnote{We set $m_{\ti u_L} = m_{\sq}$, $m_{\ti d_L} = m_{\sq} + 6\,$GeV,
$m_{\ti u_R} = m_{\sq} -19\,$GeV.} 
The production cross section of each process is shown 
in contour lines in units of pb.  
\begin{figure}[htbp]
\begin{minipage}{8cm}
\includegraphics[scale=0.9]{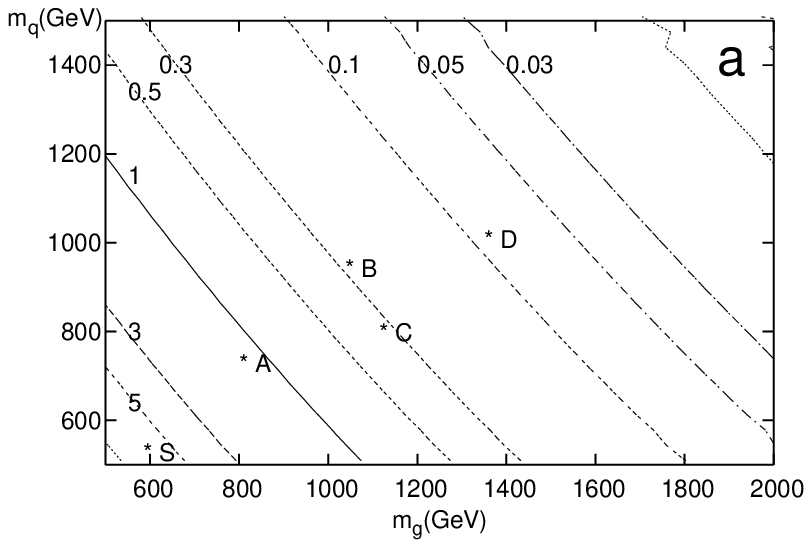}
\end{minipage}
\hfill
\begin{minipage}{8cm}
\includegraphics[scale=0.9]{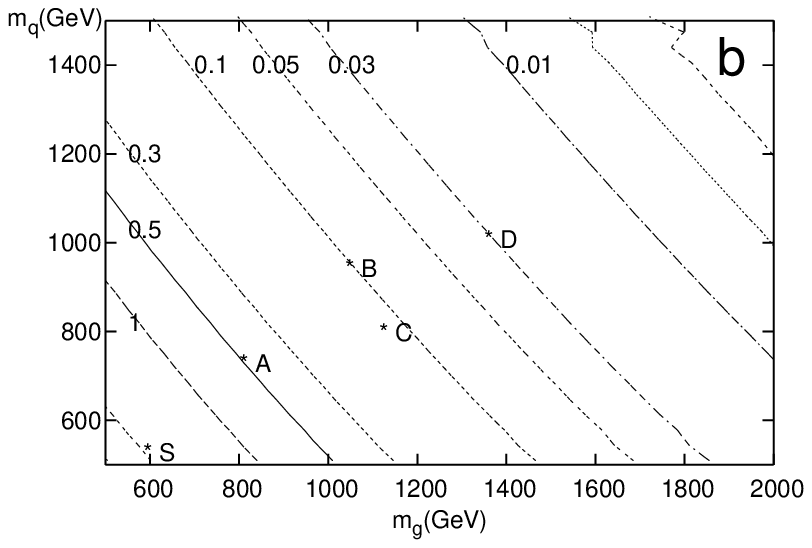}
\end{minipage}
\caption{
Contour plot of a) $\sigma(\gluino \ti u _L)$ and 
b) $\sigma(\gluino \ti d _L)$
as a function of 
$m_{\gluino}$ and $m_{\sq}$.}
\label{f2}
\end{figure}

$\sigma(\sq_L\sq_L)$ ($\sq_L = \ti u_L$ or $\ti d_L$)
are shown in Figures \ref{f1} a $\sim$ c.
$\sigma(\ti u_L \ti u_L)$ is $0.05\,$pb at $m_{\sq} = m_{\gluino} = 1000\,$GeV,
and $0.12\,$pb at $m_{\sq} = m_{\gluino} = 800\,$GeV.
This shows that they strongly depend on the squark mass $m_{\sq}$.
The mass measurement error at LHC for squarks and gluino can be around a few percent,
if the number of events is large enough \cite{LHCLC}.
$\sigma(\sq_L\sq_L)$ changes by 10\% when $m_{\sq}$ changes by 3\% around 
$m_{\sq}=m_{\gluino}=1000\,$GeV and $m_{\sq}=m_{\gluino}=800\,$GeV.
On the other hand, the cross section only weakly depends on gluino mass.
This is 
because the amplitude has a factor of $m_{\gluino}$ 
from the chirality flip,
which compensates the suppression from 
the t-channel propagator.
$\sigma(\ti u_L \ti u_L)$  changes by up to 10\% 
in the range $500\,$GeV$<m_{\gluino}<1100\,$GeV
 around $m_{\sq}=800$\,GeV,
and also  changes by up to 10\% 
in the range $700\,$GeV$<m_{\gluino}<1300\,$GeV
 around $m_{\sq}=1000$\,GeV (Figure \ref{f1}a).
In the case of $m_{\gluino}=3\,$TeV,
$\sigma(\ti u_L \ti u_L)$ is $0.04\,$pb at $m_{\sq} = 800\,$GeV,
and $0.02\,$pb at $m_{\sq} = 1000\,$GeV.
As a result, even when the accuracy of the mass measurement of the gluino
is bad, the $\sq_L\sq_L$ production cross section is a useful observable
 that can be used to quantitatively test the MSSM.
The behavior of the $\sq_R\sq_R$ production cross section 
is the same. 

$\sigma(\sq_L\sq_R)$ 
(such as $\sigma(\ti u_L\ti u_R)$ etc.) 
depends on the gluino mass more sensitively than $\sigma(\sq_L\sq_L)$.
It decreases as the gluino mass increases.
$\sigma(\ti u_L \ti u_R)$ drops by half
as $m_{\gluino}$ increases from $500\,$GeV to $1100\,$GeV
for $m_{\sq}=800\,$GeV,
and also drops by half 
as $m_{\gluino}$ increases from $700\,$GeV to $1300\,$GeV
for $m_{\sq}=1000\,$GeV (Figure \ref{f1}d).
$\sigma(\gluino\sq_L)$ 
depends on the gluino mass even more because
an onshell gluino has to be produced (Figure \ref{f2}a,\ref{f2}b).

By investigating $\sq_L\sq_L$ production processes, 
we can probe the majorana nature of gluino mass.
To measure the production cross section of 
$\ti u_L \ti u_L$ and $\ti d_L \ti d_L$,
same sign two lepton (SS2$l$) events are useful,
which is studied in \cite{Freitas:2006wd}.
This idea is as follows:
a $\ti u_L (\ti d_L)$ may dominantly produces $l^+ (l^-)$ 
through the decays of
\begin{eqnarray}
\ti u_L \to \ch^+_1 d \!\!\!&\to&\!\!\! \ti l^+ \nu_l d \to \none l^+ \nu_l d
 \ \ \ \ \ \ \ \ \ \ \ \ 
\ti d_L \to \ch^-_1 u \to \ti l^- \bar{\nu}_l u \to \none l^- \bar{\nu}_l u \cr 
 {\rm or}\ \!\!\!&\to&\!\!\! \none W^+ d \to \none l^+ \nu_l d\ ,
 \ \ \ \ \ \ \ \ \ \ \ \ \ \ \ \,
{\rm or} \ \,\to \none W^- u \to \none l^- \bar{\nu}_l u\ .
\end{eqnarray}
Therefore $l^+l^+$ events are sensitive to $\ti u_L \ti u_L$ production
and $l^-l^-$ events to $\ti d_L \ti d_L$ production.
This signature implies the existence of the Yukawa type vertex 
$g \gluino \sq$.

The ratio of $l^+l^+$ and $l^-l^-$ has more information, as studied in \cite{Baer:1995va}.
The ratio of the fractions of $u$ and $d$ in the PDF is 2:1.
Then $\sigma(\ti u_L\ti u_L)$ is larger than $\sigma(\ti d_L\ti d_L)$
and their ratio is about 4:1 (Figure \ref{f1}a,b).
Thus the ratio of the number of SS2$l$ events 
$N(l^+l^+ {\rm \ from\ } \ti u_L \ti u_L):N(l^-l^- {\rm \ from\ } \ti d_L\ti d_L)$
should be 4:1 
if leptonic branching ratios of $\ti u_L$ and $\ti d_L$ are the same.

The processes involving $\ti u_L^\ast$ and $\ti d_L^\ast$ etc. 
also become sources of SS2$l$ events
although the cross section is not large
($\sigma(\ti u_L \ti d_L^\ast) \sim 0.1\sigma(\ti u_L \ti u_L)$ 
in Figure \ref{f3}).
Since the basic observable is the sign of the leptons,
in the following we define $\sq_L^+$ as $\{\ti u_L, \ti d_L^\ast, \ti c_L, \ti s_L^\ast\}$
which can be a parent of $l^+$,
and $\sq_L^-$ as $\{\ti u_L^\ast, \ti d_L, \ti c_L^\ast, \ti s_L\}$
which can be a parent of $l^-$.

\begin{figure}[htbp]
\begin{minipage}{8cm}
\includegraphics[scale=0.9]{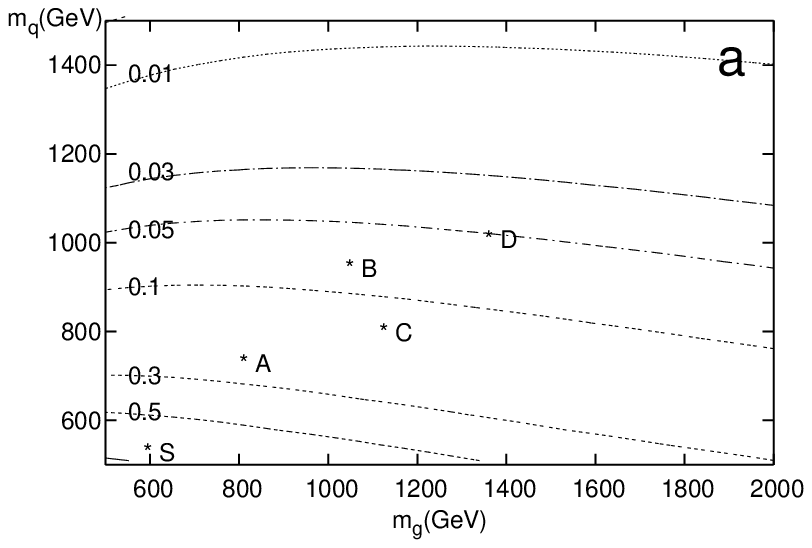}
\end{minipage}
\hfill
\begin{minipage}{8cm}
\includegraphics[scale=0.9]{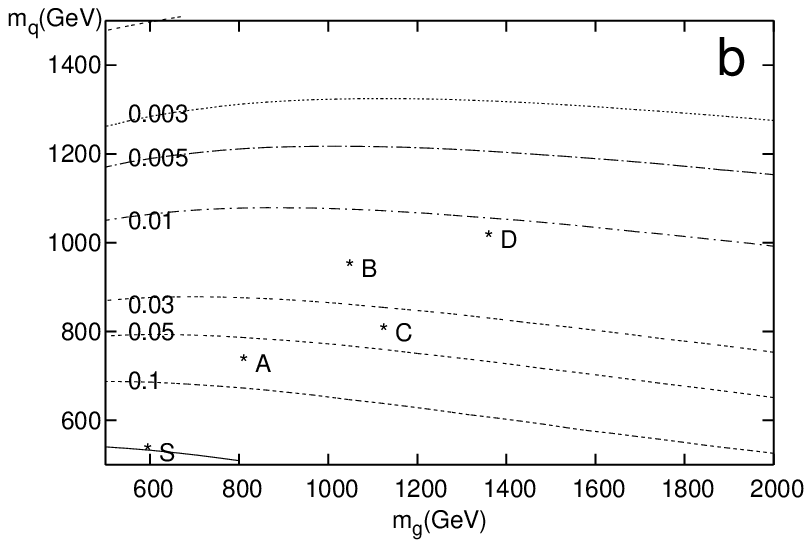}
\end{minipage}
\\
\begin{minipage}{8cm}
\includegraphics[scale=0.9]{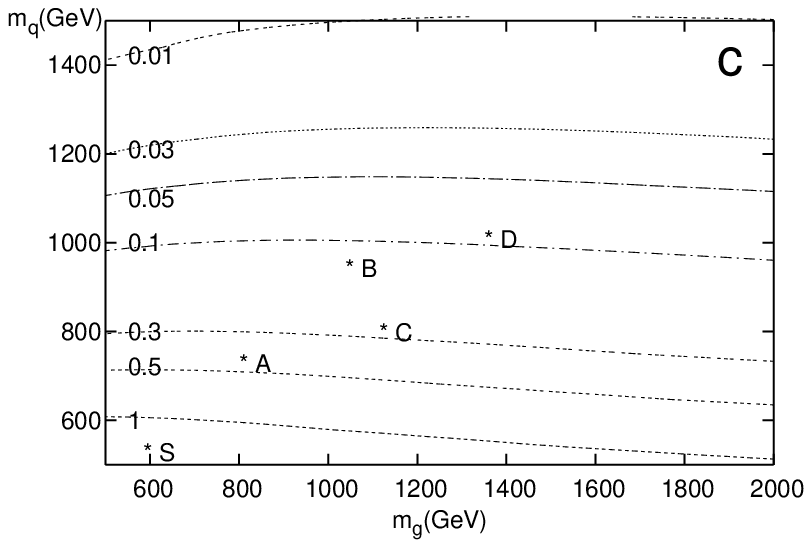}
\end{minipage}
\caption{Contour plot of a) $\sigma(\sq _L^+ \sq _L^+)$,
b) $\sigma(\sq _L^- \sq _L^-)$ and 
c) $\sigma(\sq _L^+ \sq _L^-)$ as a function of 
$m_{\gluino}$ and $m_{\sq}$.}
\label{f4}
\end{figure}

\begin{figure}[htbp]
\begin{minipage}{8cm}
\includegraphics[scale=0.9]{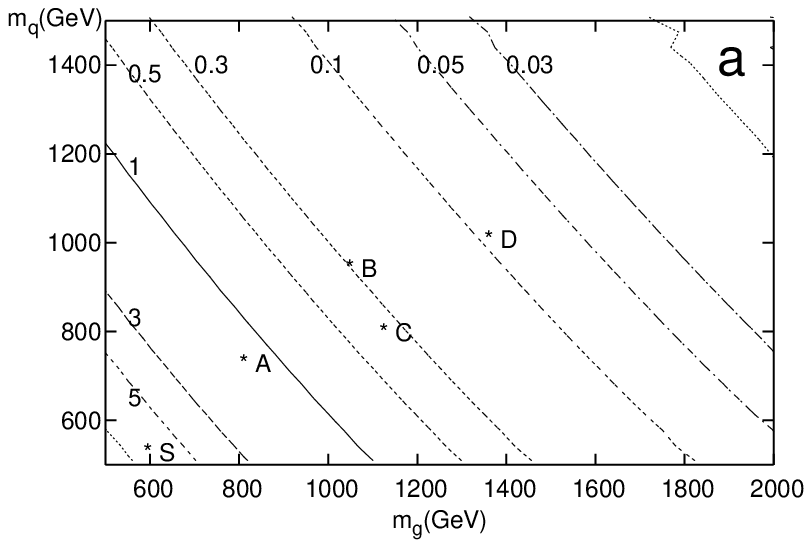}
\end{minipage}
\hfill
\begin{minipage}{8cm}
\includegraphics[scale=0.9]{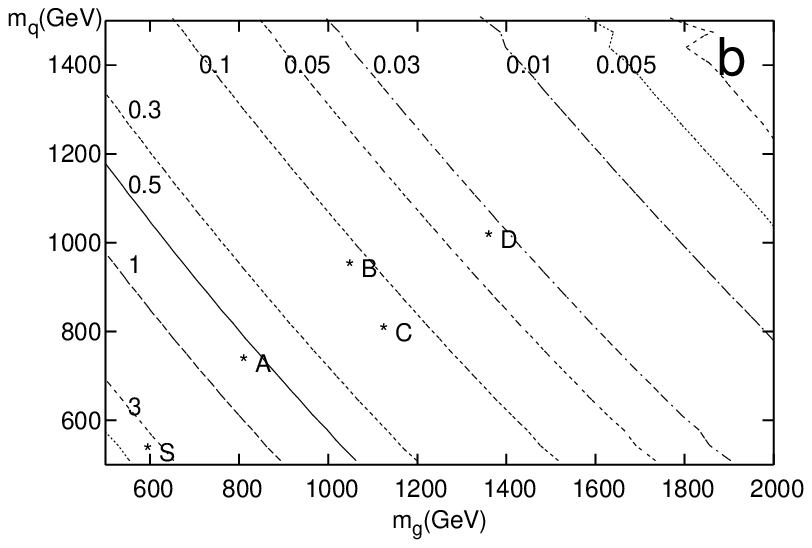}
\end{minipage}
\caption{Contour plot of a) $\sigma(\gluino \sq _L^+)$ and 
b) $\sigma(\gluino \sq _L^-)$
 as a function of 
$m_{\gluino}$ and $m_{\sq}$.}
\label{f5}
\end{figure}

\begin{figure}[htbp]
\center{
\begin{minipage}{8cm}
\includegraphics[scale=0.9]{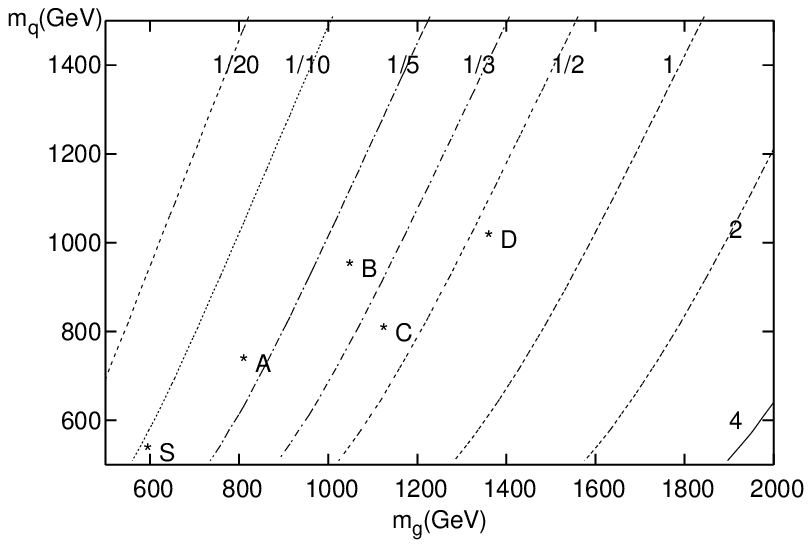}
\end{minipage}
\caption{Contour plot of the $\sigma(\sq _L^+ \sq _L^+)$ divided by $\sigma(\gluino \sq_L^+)$
as a function of $m_{\gluino}$ and $m_{\sq}$}
\label{f6}
}
\end{figure}

The mass dependencies of the production cross sections
of $\sq_L^ +$, $\sq_L^-$ 
 are shown in Figures \ref{f4}$\sim$\ref{f6}. 
By comparing Figure \ref{f4}a with Figure \ref{f1}a, 
and Figure \ref{f4}b with Figure \ref{f1}b,
we can see that $\ti u_L\ti u_L$ ($\ti d_L\ti d_L$) is dominant 
in $\sq^+_L\sq^+_L$ ($\sq^-_L\sq^-_L$) respectively.
By comparing Figure \ref{f5}a with Figure \ref{f2}a,
and Figure \ref{f5}b with Figure \ref{f2}b, 
we also find that $\gluino\ti u_L$ ($\gluino\ti d_L$),
is dominant in $\gluino\sq^+_L$ ($\gluino\sq^-_L$) respectively.

The $\gluino$ can decay into 
$\sq^\pm_L q^\mp$,
therefore $\gluino \sq^\pm_L$ production also produces $l^\pm l^\pm$.
Moreover,
$\sigma(\gluino \sq^\pm_L)$ is larger than $\sigma(\sq^\pm_L \sq^\pm_L)$
unless the gluino is too heavy (Typically $\sigma(\gluino \sq^\pm) \sim 5 \sigma(\sq^\pm \sq^\pm)$ ),
thus the $\gluino \sq^\pm_L$ production process becomes background
 to the $\sq^\pm_L\sq^\pm_L$ production process.
The ratio $\sigma(\gluino\sq^+):\sigma(\gluino\sq^-)$ is about 2:1 
(Figure \ref{f5}).
Then the ratio of the numbers of SS2$l$ events 
$N(l^+l^+ {\rm \ from\ } \gluino\sq^+):N(l^-l^- {\rm \ from\ } \gluino\sq^-)$
 should be 2:1.

The $\gluino \gluino$ production process also produces 
$l^\pm l^\pm$ and 
$N(l^+l^+ {\rm \ from\ } \gluino\gluino):N(l^-l^- {\rm \ from\ } \gluino\gluino)$
 should be 1:1.
This process, however, does not produce problematic background
because the SS2$l$ branching ratio of $\gluino\gluino$ production 
is small,
although $\sigma(\gluino \gluino)$ may be larger than 
$\sigma(\sq\sq)$ in the mSUGRA model.


The ratio of $\sigma(\sq_L^\pm \sq_L^\pm)/\sigma(\gluino\sq_L^\pm)$ 
increases as the gluino gets heavier or the squark gets lighter
(Figure \ref{f6}).
When $m_{\gluino} \sim m_{\sq}$ the ratio is always less than 1/2 
for $m_{\sq} < 1500\,$GeV.

\subsection{The MSSM model points}
The numbers of SS2$l$ events from $\gluino \sq_L^\pm$ production and 
from $\sq_L^\pm \sq_L^\pm$ production
depend on their decay patterns and the branching ratios. 
For the study in the following sections,
we choose a few model points to fix the branching ratios. 
We consider the model points with $m_{\gluino} > m_{\sq}$.
If $m_{\sq}>m_{\gluino}$, 
both of $\sq^\pm_L$ decay into $\gluino$
and this $\gluino$ can produce a lepton or an anti-lepton with the same probability,
and we cannot distinguish $\sq^\pm\sq^\pm$ and $\sq^+\sq^-$.

We take relatively heavy sparticle masses ($m_{\sq},m_{\gluino} \gsim 800$\,GeV), 
because $\sigma(\sq_L^\pm \sq_L^\pm)$ is too small 
compared with $\sigma(\gluino \sq_L^\pm)$
if the masses are lighter.
For example, at SPS1a ($m_{\sq},m_{\gluino} \sim 600$\,GeV),
which is a popular benchmark point
defined in \cite{Allanach:2002nj},
$\sigma(\gluino\sq)$ is about ten times as large as $\sigma(\sq\sq)$.
We choose four points A$\sim$D shown in Table \ref{masssss}
so that ${\sigma(\sq_L^+\sq_L^+)}/{\sigma(\gluino\sq_L^+)} \lsim 1/5$,
because we find that one can reduce $\gluino\sq$ background by factor of 1/10
by applying various cuts in Section 3.
The points A$\sim$D and SPSla are also marked in Figure \ref{f1}$\sim$\ref{f6}.
We can see in Figure \ref{f6} that
the ratio ${\sigma(\sq_L^+\sq_L^+)}/{\sigma(\gluino\sq_L^+)}$ 
is about 1/6,
 1/4,
 1/3,
 1/2,
and 1/10 for Points A$\sim$D and SPS1a respectively. 
 
Points A and B are mSUGRA points
where $(m_0,m_{\h})$ are ($100\,$GeV,$340\,$GeV)
and $(100\,$GeV, $450\,$GeV),
$A_0 = 0, \tan \beta=10, {\rm sign\,} \mu > 0$ respectively.
At these points,
the mass difference of $\ti\tau_1$ and $\none$ is small.
This feature is favored 
to reduce the LSP abundance. 
Points C and D are the mSUGRA points 
where ($m_0$,$m_{\h}$) are ($370\,$GeV, $340\,$GeV) and ($400\,$GeV, $450\,$GeV),
$A_0 = 0, \tan \beta=10, {\rm sign\,} \mu > 0$ respectively,
except the low energy gluino masses are  heavier 
than the mSUGRA predictions by 300\,GeV.
By increasing the gluino mass, 
the production cross section of the gluino decreases,
therefore $\gluino \sq$ backgrounds are smaller.
Sleptons are heavier than the lighter charginos $\ch_1^\pm$,
thus the $\ch_1^\pm$ does not decay into a slepton but decays into a $W^\pm$ boson.
These points are not favored cosmologically,
however the mass density can be reduced by tuning the mass of 
the pseudoscalar higgs boson $m_P$ as $m_P \sim 2 m_{LSP}$
without changing the rate of SS2$l$ signal
if we go beyond mSUGRA.
Moreover, the decay patterns have some similarity to 
those predicted in the LHT as we will see later.
SPS1a is also written in Table \ref{masssss} and \ref{seisei} for reference. 

The masses of some particles at our model points are
shown in Table \ref{masssss}. 
These spectra are calculated using ISAJET \cite{Baer:1993ae,Paige:2003mg}.
The other mass spectra are given in the Appendix \ref{appen1}.
The SUSY production cross sections of our model points are also shown 
in Table \ref{seisei}. 
This is calculated with HERWIG\,6.5 \cite{herwig6.5}, where we use the CTEQ 6l PDF \cite{CTEQ6l}.

\begin{table}[htbp]
\center{
\small
\begin{tabular}{|c||rrr|rrrrrrrr|}
\hline 
          &$m_0$&$m_{\h}$&$A_0$&  $m_{\gluino}$  & $m_{\ti u_L}$ & $m_{\ti u_R}$
& $m_{\ti t_1}$ & $m_{\ti b_1}$ &  $m_{\stau}$&$m_{\ti l_L}$&  $m_{\none}$         \cr
\hline 
Point A &100 &340&0& 809.86   & 737.25  & 714.56    
& 559.18 &  683.97 &  160.79 & 256.36 & 132.74\cr
Point B &100 &450&0&1047.83   &  951.16 & 919.51 
& 734.57    &  883.16  &  194.28 &324.47& 179.11 \cr
Point C &  370&340&0&  1123.23 &   808.67 & 787.79  
& 585.39 &  731.10 &  387.18 &437.13& 133.95 \cr
Point D & 400  & 450 &0&  1360.22 &  1017.91 & 988.10   
& 804.20 & 928.37 & 429.40 &503.38& 180.54 \cr
SPS1a   &100   &250 & $-100$ & 595.19    & 537.04  & 520.45 
& 379.14  & 491.92 & 133.39 &202.12    & 96.05 \cr
\hline
\end{tabular}
\caption{The mass spectra at Point A$\sim$D and SPS1a.
Here the unit of masses is GeV.}
\label{masssss}
}
\end{table}

\begin{table}[htbp]
\center{
\small
\begin{tabular}{|c||rrrrrrr|}
\hline 
     & $\sigma(SUSY)$& $\sigma(\sq^+_L\sq^+_L)$ & $\sigma(\sq^-_L\sq^-_L)$ & $\sigma(\sq^+_L\sq^-_L)$ & $\sigma(\gluino \sq^+_L)$& $\sigma(\gluino \sq^-_L)$   & $\sigma(\gluino \gluino)$ \cr
\hline 
Point A          &   8.621  &  0.2251   & 0.0674  & 0.4247 & 1.3580 & 0.6005 & 0.7134 
\cr
Point B          &   2.023  &  0.0750   & 0.0189  & 0.1309 & 0.2949 & 0.1208 & 0.1385 
\cr
Point C  &   3.418  &  0.1321   & 0.0371  & 0.2572 & 0.3410 & 0.1406 & 0.0989 
\cr
Point D &   0.963  &  0.0494   & 0.0118  & 0.0875 & 0.0897 & 0.0345 & 0.0280 
\cr
SPS1a            &  45.890  &  0.8033   & 0.2868  & $1.600\ \,$&  $7.408\ \,$& $3.544\ \,$& $4.872\ \,$ 
\cr
\hline
\end{tabular}
\caption{The SUSY production cross sections for some processes 
at Point A$\sim$D and SPS1a. Here the unit of the cross sections is pb.}
\label{seisei}
}
\end{table}

\subsection{The model with an extended gluino sector}

In this section, we consider a model with an extended gluino sector.
This model was originally considered in
\cite{Fox:2002bu,Chacko:2004mi,Carpenter:2005tz,Hisano:2006mv}
as a model with enhanced particle contents of $N=2$ SUSY 
to solve the little hierarchy problem.
The model has new fermions $\ti a^a_i$ in adjoint representations
of each SM gauge group $G_i$.
The majorana gaugino $\lambda^a_i$ has a Dirac mass term with $\ti a^a_i$.

Inspired by the model, 
we consider phenomenologies
of the gluino sector.
The mass term of the gluino is extended as follows,
\begin{equation}
-{\cal L}_{\gluino}^{\rm mass}=\frac{1}{2}m_{\gluino}\bar{\tilde{g}}\tilde{g}
\to \frac{1}{2}
\left(
\!\!\!\begin{array}{cc}
\bar{\tilde{g}} & \bar{\tilde{a}}  \\
\end{array}\!\!\!
\right)
\left(
\!\!\!\begin{array}{cc}
m_{g} & m_D \\
m_D & m_{A} \\
\end{array}\!\!\!\right)
\left(
\!\!\!\begin{array}{c}
\tilde{g}\\
\tilde{a}\\
\end{array}\!\!\!\right).
\end{equation}

Here, $\gluino$ and $\ti a$ are four component spinors 
that satisfy the majorana condition.
The $m_g$ is a majorana mass for $\gluino$,\footnote{In Ref.\cite{Chacko:2004mi}, $m_{\gluino}$ is taken as zero.}
$m_D$ is a Dirac mass between $\gluino$ and $\ti a$,
$m_A$ is a majorana mass for $\ti a$.
We leave the SU(2) and U(1) gaugino sectors unchanged. 

The mass eigenstates $\gluino_1,\gluino_2$ are majorana particles,
and the mass eigenvalues are given by
\begin{eqnarray}
m_{\gluino_{1,2}}=\frac{1}{2}\left( m_g + m_A \pm 
\sqrt{(m_g- m_A)^2 + 4m_D^2 } \right),\ \ \ \ \ \ \ |m_{\tilde{g}_1}| < |m_{\tilde{g}_2}|.
\end{eqnarray}
The mass eigenstates are defined as follows,
\begin{eqnarray}
\begin{pmatrix}
\tilde{g_1} \\
\tilde{g_2} 
\end{pmatrix}
&=&
\begin{pmatrix}
\cos \phi & \sin \phi\\
-\sin \phi & \cos \phi \\
\end{pmatrix}
\begin{pmatrix}
\tilde{g}\\
\tilde{a}\\
\end{pmatrix},\ \ \ \ \ \ \ \ \ 
(\tan \phi=\frac{ m_{\tilde{g}_1}-m_{g}}{m_D} ).
\label{phi}
\end{eqnarray}

In the limit of $m_{D} \to 0, m_A \to \infty$,
${\ti a}$ decouples from the MSSM fields, then the phenomenology becomes 
identical to that of the MSSM.
In the limit of $m_g=0,m_A=0,m_D \ne 0$
(we call this limit the pure Dirac limit),
the masses of the two gluinos become the same,
then $\gluino_1$ and $\gluino_2$ interfere strongly
so that $\gluino$ and $\ti a$ form a Dirac particle $\gluino_D$ 
and its anti particle $\bar{\gluino}_D$.
The model has continuous R-symmetry in the pure Dirac limit.
We can assign R-charge 1 for $\gluino_D$ and $\ti u_L$
,$-1$ for $\ti u_R$ and so on.
The pair production of $\ti u_L \ti u_L$, 
$\ti u_R \ti u_R$ through gluino exchange,
which is one of main production process in the MSSM at the LHC,
is forbidden by the R-charge conservation law.
This can be also understood by the fact that 
existence of nonzero majorana mass of gluino is necessary 
for $\sq_L\sq_L$, $\sq_R\sq_R$ processes.
While $\bar{\gluino}_D \ti u_L$ pair production is allowed,
$\gluino_D \ti u_L$  pair production is forbidden in the limit,
and once $\gluino_D$ is produced,
$\gluino_D$ can decay into  $\ti u_L \bar{u}$ and $\ti d_L \bar{d}$
but can not decay into $\ti u_L^\ast u$ nor $\ti d_L^\ast d$.
The difference 
between the pure Dirac limit and the MSSM is clear.
If $m_A$, $m_g \ll m_D$, the mass difference of two gluinos $\Delta m_{\gluino_{1,2}}$
is small,
continuous R-symmetry exists approximately,
and the phenomenology is similar to that in the pure Dirac limit.
We do not investigate the phenomenology of this case any further.

This model has two gluino like particles.
We may be able to observe two gluinos if the mass difference
is large enough compared with the decay widths of the gluinos
($\Gamma_{\gluino_{1,2}} \ll \Delta m_{\gluino_{1,2}} $).
In this case, each gluino decay produces $\sq^+$ and $\sq^-$
with the same branching ratio.
If $\gluino_2$ is too heavy
so that the production cross section is too small to be observed at LHC,
we can observe only $\gluino_1$. 
It is not possible in this case to distinguish this model from the MSSM
only by the mass spectrum.
We focus on this case and study the deviation of 
the production cross sections from the MSSM predictions. 
The gluino sector of this model has two degrees of freedom in addition to $m_{\gluino_1}$
and we take them as the other gluino like particle's mass $m_{\gluino_2}$
and the majorana mass $m_g$.
 
\begin{figure}[htb]
\center{
\begin{minipage}{7.7cm}
\includegraphics[scale=0.6]{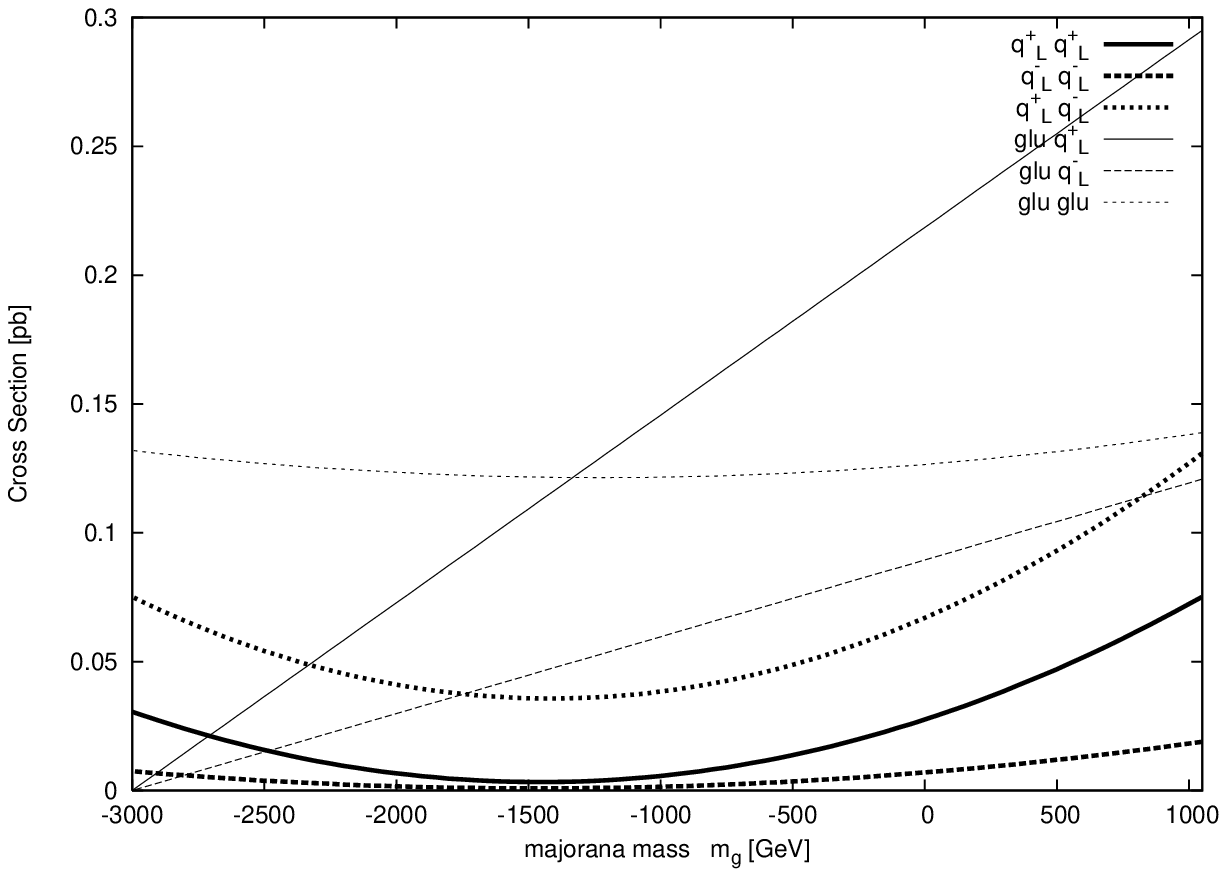}
\caption{The production cross sections 
as a function of the gluino majorana mass $m_g$
for the model with an extended gluino sector.
The mass spectrum is the same as that at Point B.}
\label{diracB}
\end{minipage}
\hfill
\begin{minipage}{7.7cm}
\includegraphics[scale=0.6]{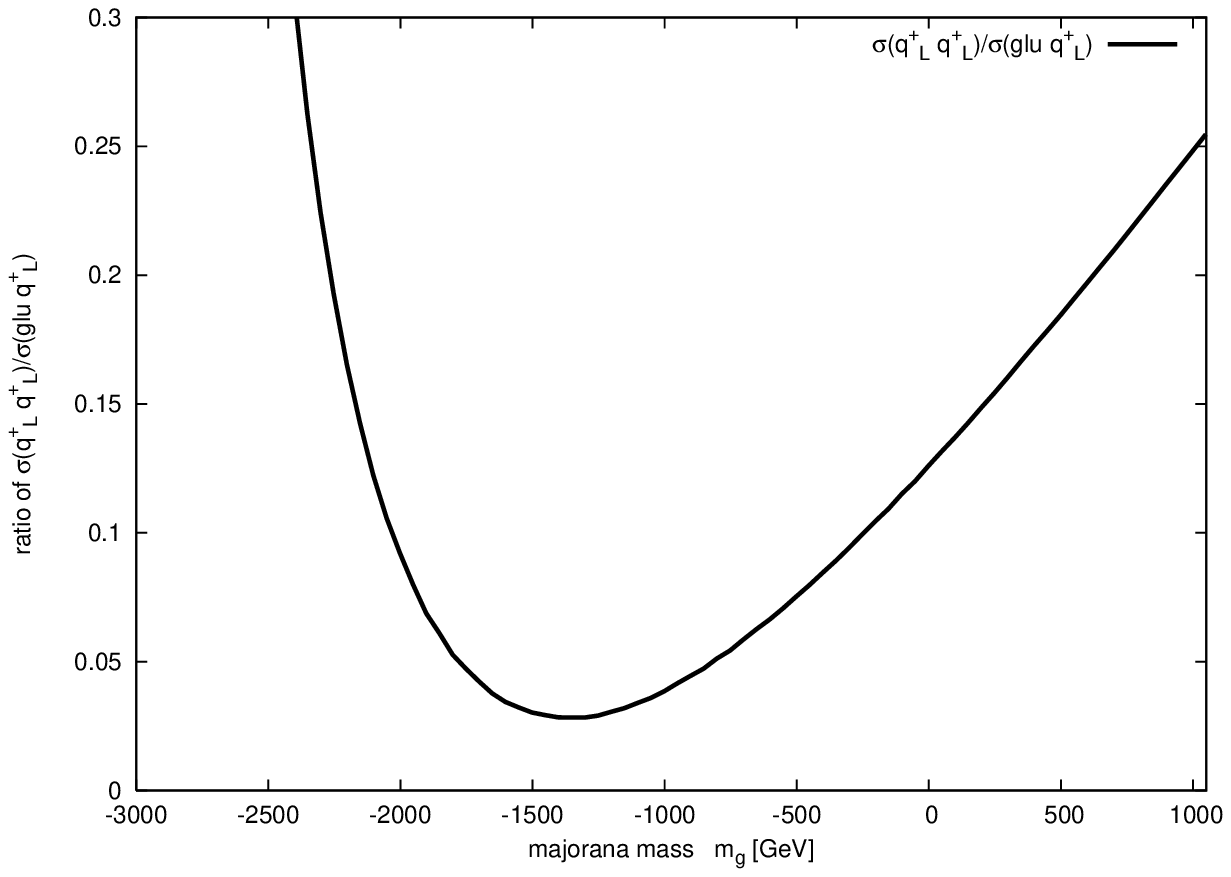}
\caption{The ratio of $\sigma(\sq_L^+ \sq_L^+)$
and $\sigma(\gluino \sq_L^+)$ 
as a function of the gluino majorana mass $m_g$
for the model with an extended gluino sector.
The mass spectrum is the same as that at Point B.}
\label{PointBratio}
\end{minipage}
}
\end{figure}

In Figure \ref{diracB}, we show the production cross sections 
of SUSY processes such as $\tilde{q}_L^\pm\tilde{q}_L^\pm$ and
 $\gluino\tilde{q}_L^\pm$ as functions of the
 majorana gluino mass $m_g$.
Here, we fix $m_{\gluino_2} =-3000\,$GeV
and $m_{\gluino_1} =1047.83\,$GeV,
which is the gluino mass at Point B.
The $\gluino_2$ cannot be searched for at LHC.
We take the mass spectrum of the other sparticles 
to be the same as that of Point B.

In the limit of $m_{g} = m_{\gluino_1}= 1047.83\,$GeV,
$m_D$ is zero and $m_{\gluino_2} = m_A= -3000\,$GeV.
The $\gluino_2$ decouples from $\sq$,
and the cross sections involving $\gluino_1$ are the same as those of Point B. 
Changing $m_g$ from this value
distorts the model from the MSSM.
As the majorana mass $m_g$ decreases 
keeping $m_{\gluino_1}$ and $m_{\gluino_2}$ fixed,
the total SUSY production cross section decreases.
In particular,
$\sigma(\sq_L^\pm \sq_L^\pm)$ decreases quickly,
while  $\sigma(\gluino_1\sq^\pm)$ decreases linearly
because $\sigma(\gluino_1\sq^\pm)$ is proportional to $\cos^2\!\phi$ and 
$m_g=\cos^2\!\phi (m_{\gluino_1} -m_{\gluino_2}) + m_{\gluino_2}$.  
The fraction of $\gluino$ in $\gluino_2$ increases,
 but $\gluino_2$ is too heavy
so that $\sigma(\gluino_2 \sq)$ is small.
$\sigma(\gluino_1\gluino_1)$ is less sensitive to the majorana mass $m_g$.
This means that the squark exchange diagram does not 
contribute much to the cross section.

When the majorana mass $m_g=0$,
$\sigma(\sq_L^\pm \sq_L^\pm)$ is still nonzero.
This is because the majorana mass of the adjoint fermion $m_A$ causes the chirality flip.
There are also minor contributions from $\sigma(\ti u_L \ti d_L^\ast)$
and so on,
which are not suppressed by the $m_g$ factor.

As $m_g$ decreases further (absolute value $|m_g|$ increases),
$\sigma(\sq_L^\pm\sq_L^\pm)$ approaches zero around $m_g \sim -1500\,$GeV,
while $\sigma(\gluino_1\sq_L^\pm)$ is reduced by factor of 3.
This behavior can be explained 
by the  dependence of the subprocess cross section $\sigma(qq \to \sq_L \sq_L)$
on the mass parameters as follows,
\begin{eqnarray}
&&\!\!\!\!\!\!\!\!\!\!\sigma(q q \to \sq_L \sq_L) =
\frac{\beta_{f}}{64\pi s} \int_{-1}^{1} d(\cos\theta) 
\frac{4E^2 g_s^4}{9}\cr
&&\ \ \ \ \ \ \ \ \ \ \ \ \ \ \ \ \ 
\times
\left|
T^a_{ij}T^a_{kl}\{f(m_{\gluino_1},m_{\sq},\mathbf{p})\cos^2\!\phi+f(m_{\gluino_2},m_{\sq},\mathbf{p})\sin^2\!\phi \}
\right.
\cr
&&\left.
\ \ \ \ \ \ \ \ \ \ \ \ \ \ \ \ \ \ \ \ \ \ \ \ \ \ \ \ \ \ \ + 
T^a_{il}T^a_{kj}\{f(m_{\gluino_1},m_{\sq},-\mathbf{p})\cos^2\!\phi+f(m_{\gluino_2},m_{\sq},-\mathbf{p})\sin^2\!\phi \}
\right|^2
\cr
&&\ \ \ \ \ \ \ \ \ \ \ \ \,\propsim
\left|
\frac{m_{\gluino_1}\cos^2\phi}{m_{\gluino_1}^2 +m_{\sq}^2}
+ \frac{m_{\gluino_2}\sin^2\phi}{m_{\gluino_2}^2 +m_{\sq}^2}
\right|^2, 
\label{appp}
\end{eqnarray}
\begin{eqnarray}
{\rm where,\ }f(m_i, m_f, \mathbf{p}) 
\equiv \frac{m_{i}}{m_{i}^2 + m_f^2 + 2 |\mathbf{p}|^2 - 2 E |\mathbf{p}|\cos \theta}
\simeq \frac{m_{i}}{m_{i}^2 + m_f^2}.
\label{appp1}
\end{eqnarray}
Here $\sigma(qq \to \sq_L\sq_L)$ is the cross section of the subprocess 
$qq \to \sq_L\sq_L$ with center of mass energy $s$. 
$\mathbf{p}$ is the momentum of one of the created $\sq_L$.
We set the z-axis along the momentum of one of the initial quarks,
 $\theta$ is the polar angle of $\mathbf{p}$ from the z-axis,
 and $\beta_f$ is the beta factor ($\beta_f = \sqrt{1 - 4m_{\sq}^2/s}$).
$T^a$ denotes a generator of the SU(3) group
and $g_s$ denotes the gauge coupling.
$\phi$ is the mixing angle between the $\gluino$ and $\ti a$
as defined in (\ref{phi}).
In the last line in eq.(\ref{appp}) and the last equality in eq.(\ref{appp1}),
we take the limit that $|\mathbf{p}| \ll m_{\sq}$ 
because $\sigma(pp \to \sq_L\sq_L)$ 
is dominated by the threshold production. 
We can calculate that
$\sigma(\sq_L\sq_L)$ approaches zero around $\phi = \phi_0$
defined as
\begin{eqnarray}
\tan\phi_0 =\sqrt{- \frac{m_{\gluino_1}}{m_{\gluino_2}}
\frac{m_{\gluino_2}^2 +m_{\sq}^2}{m_{\gluino_1}^2 +m_{\sq}^2}}.
\end{eqnarray}
This corresponds to $m_{g} = -1516$\,GeV for Point B
(here, we use the relation 
$m_{g} = \cos^2\!\phi\, m_{\gluino_1} + \sin^2\!\phi\, m_{\gluino_2}$).
Note that we again neglect the contributions 
from chargino and neutralino exchange diagrams in the calculation.
They are at most of the order of $10^{-3}$ pb,
and negligible.

$\sigma(\sq^+_L \sq^+_L)$/$\sigma(\gluino \sq^+_L)$
is less than $5\%$
in the range of $-1800\,{\rm GeV} <m_g<-800\,$GeV,
while $\sigma(\sq^+_L \sq^+_L)/\sigma(\gluino \sq^+_L)$
is about $25\%$ in the MSSM limit ($m_g=1047.83\,$GeV).
The production cross section $\sigma(\sq_L\sq_L)$ is reduced 
by more than factor of 5
compared with $\gluino\sq_L$ in this range (Figure \ref{PointBratio}).

As $m_g$ decreases further,
$\sigma(\sq_L^\pm\sq_L^\pm)$ increases again 
while $\sigma(\gluino_1\sq_L^\pm)$ keeps decreasing.
When $m_g= -3000$\,GeV, $\gluino_2$ is $\gluino$ and $\gluino_1$ is $\ti a$.
In this limit, $\sigma(\sq_L\sq_L)$ is enhanced by the factor of $m_g$
in the amplitude,
and $\sigma(\gluino_1 \sq_L^\pm)=0$ 
because $\gluino_1$ does not couple to $\sq$.
We do not discuss this region because the production and 
decay pattern would be significantly different from the MSSM.

Note that
the mass spectrum is the same as that at Point B
for entire $m_g$ in Figure \ref{diracB}. 
We can detect the deviation from the MSSM 
only through measurements of production cross sections 
such as $\sigma(\sq_L^\pm\sq_L^\pm)$, $\sigma(\gluino\sq_L^\pm)$
or their ratios.

\subsection{The Littlest Higgs model with T-parity}
The Littlest Higgs model with T-parity (LHT) 
\cite{Cheng:2003ju,Cheng:2004yc,Hubisz:2004ft} is an alternative scenario that solves
the quadratic divergence problem for Higgs mass
and predicts a stable DM candidate.
This model is the extension of the Littlest Higgs model 
\cite{Arkani-Hamed:2002qy}.
The features of the Littlest Higgs model are:
\begin{enumerate}
\item Higgs bosons are introduced as pseudo Nambu-Goldstone bosons
of global symmetry breaking $SU(5)/SO(5)$.
The global symmetry is partially gauged,
and the gauge symmetry is $[SU(2)\times U(1)]^2$.

\item The gauge groups are spontaneously broken at scale $f$ as 
$[SU(2)\times U(1)]^2 \to SU(2)_L\times U(1)_Y $.
Heavy gauge bosons of the broken gauge groups are called
as $W_H^\pm$, $Z_H$, $A_H$.
The masses are,
\begin{eqnarray}
M_{Z_H}\sim M_{W_H} \simeq g f \left[ 1- \frac{v^2}{8f^2} \right],\ \ \ 
M_{A_H} \simeq \frac{g^{\prime}f}{\sqrt{5}} \left[ 1-\frac{5v^2}{8f^2} \right].
\end{eqnarray}
Here, $v$ is the electroweak symmetry breaking scale, $v \simeq 246\,$GeV.

\item To cancel the quadratic divergence of Higgs mass,
the third generation fermion sector has to be extended 
to respect the global symmetry of the theory.
In particular, $T_+$ has to be introduced as a partner of the $t$ quark.
The $T_+$ is a SU(2)$_L$ singlet Dirac fermion.
\end{enumerate}
However, 
this model suffers from large tree level corrections 
to the electroweak parameters.
Even if the parameters of the models are tuned to reduce the corrections,
$f$ becomes large enough that the fine tuning problem is reintroduced \cite{LHfinetuning,Hewett:2002px}.
To solve this problem,
the LHT model imposes invariance under the T-parity that corresponds
to switching the two [SU(2)$\times$U(1)] gauge groups.
Matter sectors are extended 
so that there is a T-odd partner for each SM fermion.

The T-parity plays a similar role to the R-parity of the SUSY model.
The Lightest T-parity odd particle (LTP) cannot decay 
because T-parity is multiplicatively conserved for all vertices.
The LTP can be a candidate for DM.
Moreover, T-odd particles can be produced only in pairs
 in a collider experiment and 
each of them must decay into final states including an
odd number of T-odd particles.
As a result, the final states include at least two LTPs.
The collider signal at LHC is 
large transverse missing energy $E\slush_T$, just like SUSY.

This model predicts a set of new particles.
Amongst them,
the heavy gauge bosons $W_H^\pm,Z_H,A_H$,
the SM fermion partners $u_-,d_-,s_-,c_-,b_-,t_-$,
and the top partner $T_-$ are    
 T-odd.
In the following,
$q_-^+$ denotes \{$u_-,c_-,t_-,\bar{d}_-,\bar{s}_-,\bar{b}_-$\},
$q_-^-$ denotes \{$d_-,s_-,b_-,\bar{u}_-,\bar{c}_-,\bar{t}_-$\},
$q_-$ denotes $q_-^+$ and $q_-^-$.

No T-odd partner for the SU(3) gauge boson is introduced. 
On the other hand, 
the decay pattern of T-odd $q_-$ is similar to that of $\sq$ in SUSY.
According to Ref.\cite{Belyaev:2006jh},
about 60\% of $q^+_-$ decays into $W^+_H$,
100\% of $W^+_H$ decays into $W^+$
and 25\% of $W^+$ decays into leptons.
Therefore about 15\% of $q^+_-$ decay leptonically.
This decay pattern is similar to that of $\sq_L$ at Point C.
Thus, for the LHT model,
we obtain SUSY-like signal 
as if there are only $\sq\sq$ and $\sq\sq^\ast$ production at LHC.
Although there is no t-channel colored particle exchange,
$\sigma(q_- q_-)$ is non zero due to the t-channel exchange of $W_H$ and $A_H$.
It is higher than that of the MSSM because $q_-$'s are fermions.
For example, $\sigma(q^+_-q^+_-)$ is 0.7\,pb, $\sigma(q^-_-q^-_-)$ is 0.15\,pb
 at $M_{q_-} = 800\,{\rm GeV}$ and $f=560\,$GeV,
and $\sigma(q^+_-q^+_-)$ is 0.2\,pb, $\sigma(q^-_-q^-_-)$ is 0.04\,pb
at $M_{q_-} = 1000\,{\rm GeV}$ and $f=700\,$GeV.
$\sigma(q^+_-q^+_-)$ is $4\sim 5$ times as large as $\sigma(\sq^+\sq^+)$ at
the MSSM model points with the same mass scale (See Figure \ref{f4}a).
The ratio $\sigma(q_-^+q_-^+)/\sigma(q_-^- q_-^-)$ is similar to 
$\sigma(\sq^+ \sq^+)/\sigma(\sq^- \sq^-)$ of the MSSM.
The ratio $\sigma(l^+ l^+)/\sigma(l^- l^-)$ of the LHT is expected
to be higher than that of the MSSM, because $\gluino\sq$ production dominates
the total SUSY production.
As we will see later,
the production cross sections and their ratios will help to distinguish 
the LHT and the MSSM.

\subsection{Summary of the production cross sections}

There are several models which predict a MSSM-like collider signature
with large $E\slush_T$.

In section 2.2 
we have shown that the 
$\sq_L\sq_L$
production cross section changes significantly
if the majorana mass contribution to the gluino mass is reduced
in the model with an extended gluino sector.
In particular, in the case of $m_g<m_{\gluino_1}$,
$\sigma(\sq_L\sq_L)$ can be reduced significantly
compared to $\sigma(\gluino\sq_L)$.

On the other hand,
for the LHT model,
$q_-$ may have a similar decay pattern to $\sq$
while there is no particle corresponding to gluino.
Hence,
the signal is similar than that of the MSSM with an undetectably heavy gluino.
$\sigma(q_-q_-)$ is
larger than $\sigma(\sq\sq)$ by a factor of 5.

The signal cross sections of these models are different to those of the MSSM,
even if the mass spectrum is the same.
Therefore $\sigma(\sq_L\sq_L \to l^\pm l^\pm + X)$ is 
one of the key observables for MSSM studies.

\section{Separation of $\sq\sq$ and $\gluino\sq$ productions}
\subsection{Branching ratios of $\sq$, $\gluino$}
To identify
$\sq_L\sq_L$ production 
(mainly $\ti u_L\ti u_L$ production),
$\sq_L \sq_L \to 2l + X$ events would be useful because
${\cal BR}(\sq_L^\pm \to l^\pm + X)$ $\gg$ ${\cal BR}(\sq_L^\pm \to l^\mp +X)$.
In this paper, we assume 100\% of $\sq_R$ decays into q and the LSP $\none$.
This is indeed realized over most of the mSUGRA parameter space.
However, in less constrained models neutralino mixing can be different,
such that also $\sq_R$ has cascade decays involving 
the heavier neutralinos and charginos.
In this case, $\sq_L\sq_R$ and $\sq_R\sq_R$ production can both produce SS2$l$ events,
and the following analyses are more complicated.
In particular, $\sigma(\sq_L\sq_R)$ is not so sensitive 
to the majorana nature of gluino, 
therefore it is more difficult to probe the majorana nature of the gluino 
using the SS2$l$ channel.
We assume that the LSP $\none$ is dominantly a Bino 
and the $\ntwo$ a Wino,
so $\sq_R$ does not produce leptons.

The signal rate depends on the leptonic branching ratios of sparticles.
The branching ratios at Points A$\sim$D are shown in Appendix \ref{appbranch}.
We summarize them in Table \ref{lepbranch}.
\begin{table}[htbp]
\small
\begin{minipage}{8.8cm}
\small
\begin{tabular}{|cl||r|r|r|r|}
\hline
 \multicolumn{2}{|c||}{mode} & \multicolumn{4}{c|}{BR(\%)} \\
\cline{3-6}
 &  & \multicolumn{1}{c|}{Point A} & \multicolumn{1}{c|}{Point B} & \multicolumn{1}{c|}{Point C} & \multicolumn{1}{c|}{Point D}\\
\hline
\hline
$ \gluino    $ & $ \to \sq_L^+ q$ & $11$ & $10$ & $14$ & $13$  \\
$            $ & $ \to \sq_L^- q$ & $11$ & $10$ & $14$ & $13$ \\
$            $ & $ \to \sq_R q$ & $38$ & $36$ & $32$ & $32$ \\
$            $ & $ \to \sq_3 q_3$ & $41$ & $45$ & $40$ & $41$  \\
\hline
$ \ti u_L $ & $ \to l^+ X   $ & $31$ & $46$ & $18$ & $17$  \\
$ \ti d_L $ & $ \to l^- X   $ & $30$ & $44$ & $17$ & $17$  \\
\hline
$ \ti b_1 $ & $ \to l^- X   $ & $28$ & $39$ & $17$ & $20$  \\
$ \ti b_2 $ & $ \to l^- X   $ & $22$ & $27$ & $11$ & $12$  \\
$ \ti t_1 $ & $ \to l^+ X   $ & $29$ & $37$ & $17$ & $10$  \\
$ \ti t_2 $ & $ \to l^+ X   $ & $18$ & $23$ & $6.3$ & $11$  \\
\hline
\end{tabular}
\end{minipage}
\begin{minipage}{7.5cm}
\caption{Branching ratios of squarks and gluinos at Point A$\sim$D.
These are calculated by ISAJET.
Here, $\sq_L^+$ denotes
\{$\ti u_L,\ti d_L^\ast,\ti c_L,\ti s_L^\ast$\},
$\sq_L^-$ denotes 
the antiparticles of $\sq_L^+$,
$\sq_R$ denotes \{$\ti u_R, \ti d_R, \ti c_R, \ti s_R$\} 
and their antiparticles.
$\sq_3$ denotes \{$\ti b_1,\ti b_2,\ti t_1,\ti t_2$\} 
and their antiparticles.
$X$ means LSP and other SM particles. 
}
\label{lepbranch}
\end{minipage}
\end{table}

We can see that
$\sq_L^\pm$ produces $l^\pm$
when $\sq_L^\pm$ decays through a chargino.
${\cal BR}(\sq_L^\pm \to l^\pm )$
is about 30\% at Point A.
If $\sq_L^\pm$ decays through a neutralino,
$\sq_L^\pm$ may also produce $l^\mp$.

On the other hand, 
a gluino decays into third generation squarks $\sq_3$ more than
the 1st and 2nd generation squarks $\sq_L$,
and also decays into $\sq_R$ more than $\sq_L$ at these points,
because gluino and squark masses are close and 
the phase space of the gluino decay is sensitive 
to the small differences of squark masses.
(The third generation squarks $\ti t,\ti b$ are lighter than
the 1st, 2nd generation $\sq_L$,
and $\sq_R$ is lighter than $\sq_L$.)
We also find that
${\cal BR}(\gluino \to l^+ ) = {\cal BR}(\gluino \to l^- )$
and they are small (8\%),
because $\gluino$ decays dominantly into $\sq_R$.
As a result,
${\cal BR}(\sq_L^+\sq_L^+ \to l^+l^+)$
is 9\%,
${\cal BR}(\gluino\sq_L^+ \to l^+l^+)$ is
2.4\%,
${\cal BR}(\gluino\gluino \to l^+l^+)$ is 
0.6\% at Point A.

To measure $\sigma(\sq_L^+\sq_L^+)$,
we need to reduce the events involving $\gluino$ by means of appropriate cuts.
We can achieve this in part by rejecting events with $b$-tagged jets
because ${\cal BR}(\gluino \to \ti t$ or $\ti b )$ is large;
${\cal BR}(\gluino \to l$ without $b$-quark)
is only 6\%.
For pair production processes,
${\cal BR}(\sq_L^+\sq_L^+ \to l^+l^+$ without $b$-quark)
is $\sim$ 9\%,
${\cal BR}(\gluino\sq_L^+ \to l^+l^+$ without $b$-quark)
$\sim$ 1\%,
${\cal BR}(\gluino\gluino \to l^+l^+$ without $b$-quark)
$\sim$ 0.1\%.
However, the efficiency of $b$-veto is at most 60\%.

In the following, we neglect SS2$l$ events from $\gluino \gluino$ production.
This is reasonable if squark and gluino masses are sufficiently large,
 such as at Points A $\sim$ D,
 because $\sigma(\gluino\gluino)$ is small (Table \ref{seisei})
and ${\cal BR} (\gluino \to \sq_L) \ll 1$.
Furthermore, the events contain more $b$-jets on average.
 
\subsection{Event generation and detector simulation}
We generate about 300,000 SUSY events 
using HERWIG\,6.5 \cite{herwig6.5} at Points A$\sim$D and SPS1a.
The number of events actually produced by HERWIG and 
the corresponding integrated luminosities
are listed in Table \ref{gene}.
$N(\sq_L^+\sq_L^+):N(\sq_L^-\sq_L^-)$ is about 4:1 
and $N(\gluino\sq_L^+):N(\gluino\sq_L^-)$ is about 2:1 
for these model points as discussed in Sec \ref{MSSMprod}.
\begin{table}[htbp]
\small
\center{
\begin{tabular}{|c||rrrrrrr||r|}
\hline 
     & $N(SUSY)$ &$ N(\sq^+_L\sq^+_L) $ & $ N(\sq^+_L \sq^-_L )$ & $N( \sq^-_L\sq^-_L) $ & $ N(\gluino \sq^+_L )$ &$ N(\gluino \sq^-_L) $   & $N( \gluino \gluino) $ &  $\int\!\! dt {\cal L} $\cr
\hline 
Point A  &  289906 &  7865  & 10698 &  2197 & 44007 & 19330  & 21065 &  33.63\cr
Point B  &  284544 & 10601  & 11115 &  2526 & 39208 & 15818  & 13242 & 140.65\cr
Point C  &  295042 & 11411  & 14666 &  3072 & 25793 & 10729  &  4748 &  86.32\cr
Point D  &  295695 & 14505  & 14394 &  3279 & 23589 &  8982  &  3333 & 307.06\cr
SPS1a    &  293161 &  5412  & 10423 &  1849 & 46072 & 22241  & 31371 &  6.39 \cr
\hline
\end{tabular}
\caption{Numbers of events generated by HERWIG\,6.5.
Here, the unit of the integrated luminosities is fb$^{-1}$.}
\label{gene}}
\end{table}

We use AcerDET \cite{Richter-Was:2002ch} for event reconstruction.
AcerDET is a fast simulation and reconstruction package.
It finds jets, isolated electrons, muons, photons
and calculates the missing transverse energy from particles in the events. 
The granularity of the calorimetric cells is assumed as 
($0.1\times0.1$) in ($\eta\times\phi$) coordinates
for $|\eta|<3.2$.
The clusters with $p_T > 15$\,GeV for $\Delta R_{\rm cone} =0.4$ 
are classified as jets.
It also labels a jet as a $b$-jet if a $b$-quark with
momentum $p_T > 5$\,GeV is found within the cone $\Delta R = 0.2$.
The tagging efficiency of the algorithm is about 80\% 
and it is too high compared with the full simulation result of 60\% in ATLAS.
Therefore we assume that 60\% of the $b$-labeled jets are tagged.
Isolation criteria for muons, electrons and photons are 
$p_T >10$\,GeV and $|\eta| < 2.5$,
separation by $\Delta R >0.4$ from other clusters 
and $\sum E_T < 10$\,GeV in a cone $\Delta R <0.2$ around them.
For electrons and photons, we require 
$\Delta R_{\rm cluster} < 0.1$.
The $E\slush_T$ is defined as 
\begin{equation}
E\slush_T = \left| \sum_{{\rm visible}}\mathbf{p}_T \right|,
\end{equation}

and
calculated by summing the transverse momenta of 
all cells as follows,
\begin{equation}
E\slush_T = \left|\sum_{{\rm cells}}\mathbf{p}_T\right|.
\end{equation}

The numbers of SS2$l$ events from each production process 
are shown in Table \ref{bprod} for Point B.
\begin{table}[htbp]
\center{
\small
\begin{tabular}{|c||r|rrr|rrr|}
\hline 
{Point B} & generated &\multicolumn{3}{l|}{$l^+l^+$} & \multicolumn{3}{l|}{$l^-l^-$ } \cr
\cline{3-8}
&  & all & $c_0$ & $c_1$  & all & $c_0$ & $c_1$ \cr
\hline
\hline 
 total               & 284544&  4363&  2573&  1680 & 2231&  1288&  749\cr
\hline
 $\sq_L^+\sq_L^+$    &  10601&  1410&   967&   958 &    6&     2&    2\cr
 $\sq_L^-\sq_L^-$    &   2526&     1&     1&     1 &  399&   264&  258\cr
 $\sq_L^+\sq_L^-$    &  11115&    88&    54&    52 &  112&    70&   68\cr
 $\gluino \sq_L^+$   &  39208&  1720&  1067&   467 &  149&    84&   31\cr
 $\gluino \sq_L^-$   &  15818&    46&    31&    14 &  732&   469&  235\cr
 $\gluino\gluino$    &  13242&   220&   117&    22 &  225&   121&   26\cr
\hline 					
\end{tabular}
\caption{Numbers of SS2$l$ events from each production process at Point B}
\label{bprod}}
\end{table}

Here, all, $c_0$ and $c_1$ denote 
the different cuts applied to the events,
\begin{eqnarray}
{\rm all}&:& {\rm all\  SS2}l{\rm \  events\ generated\ by\ HERWIG\,6.5.}
\cr
c_0&:& {\rm basic\ cuts,}\ E\slush_T > 200\,{\rm GeV,}\ M_{\rm eff} > 500 \,{\rm GeV}
,\ E\slush_T > 0.2 M_{\rm eff} {\rm \ and\ } n_{100}\ge 2
\cr&&
(n_{100} {\rm \ is\ the\ number\ of\ jets\ with\ }
 p_T \ge 100\,{\rm GeV})
\cr
c_1&:& c_0{\rm \ and}\ n_b =0.\ 
(n_b {\rm \ is\ the\ number\ of\ }b\verb|-|{\rm tagged\ jets.})
\nonumber
\end{eqnarray}
$M_{\rm eff}$ is defined as
\begin{eqnarray}
M_{\rm eff}= \sum_{\scriptscriptstyle {\rm jets} \atop \scriptscriptstyle |\mathbf{p}_T|\ge 50\,{\rm GeV}}
\!\!\!\!\!|\mathbf{p}_T|
\ \ +
\sum_{\scriptscriptstyle {\rm leptons} \atop \scriptscriptstyle |\mathbf{p}_T|\ge 10\,{\rm GeV}}
\!\!\!\!\!|\mathbf{p}_T| \ \ +\ \  E\slush_T.
\end{eqnarray}

The SS2$l$ events are mainly produced by $\sq_L^\pm\sq_L^\pm$, $\gluino\sq_L^\pm$, $\gluino\gluino$.\footnote{The other SS2$l$ events come mainly from productions involving third generation squarks, charginos and gluino.}
We find that $N(l^+l^+$ from $\sq^+_L \sq_L^+$):$N(l^-l^-$ from $\sq^-_L \sq_L^-$)
is nearly 4:1 and 
$N(l^+l^+$ from $\gluino \sq_L^+$):$N(l^-l^-$ from $\gluino \sq_L^-$)
is nearly 2:1.
Note that $l^\pm l^\pm$ events are also produced from $\sq^+\sq^-$,
if $\sq$ decays into a neutralino and the neutralino decays into $\tau^+\tau^-$.
After the $c_0$ cut,
contributions from $\sq^\pm_L\sq^\pm_L$,$\sq^+_L\sq^-_L$,
$\gluino\sq^\pm_L$,$\gluino\gluino$ dominate the SS2$l$ events.
We describe the number of generated $l^\pm l^\pm$
events after cut $c_i$ as $N(l^\pm l^\pm;{c_i})$.
Comparing $N(l^\pm l^\pm;c_0)$ with $N(l^\pm l^\pm;c_1)$,
we see that the $b$-veto cut ($c_1$) reduces only events involving $\gluino$.
The $l^+l^+$ events from $\gluino \sq^\pm$ are reduced by half.
The $l^+l^+$ events from $\gluino \gluino$ are reduced by one fifth.
After the $c_1$ cut, $l^+l^+$ events from $\sq^+\sq^+$ are 60\% of all $l^+l^+$ events,
while they are 40\% under the $c_0$ cut.

\subsection{Hemisphere cuts}
To further reduce events involving $\gluino$,
we next study the number of jets emitted from $\sq$ and $\gluino$.
A gluino decays into a squark and a quark,
and the squark decays into a chargino or a neutralino and a quark.
Thus, a gluino usually emits at least two jets
while a squark emits at least one jet.
We can distinguish the parent particles
by the number of high $p_T$ jets in the events.
In this paper,
we divide the final state particles into two groups
called hemispheres,
then we investigate the number of jets and the invariant masses in each hemisphere.

SUSY production processes always occur in pairs due to R-parity conservation.
Particles from each sparticle decay with momentum $p_i,p_j,...$ 
are kinematically constrained
so that $(p_i + p_j + ...)^2 = m^2$ 
where $m$ is the mass of the parent sparticle.
When the parent sparticle is boosted
the decay products are boosted in the same direction.
We therefore divide all high $p_T$ objects into two groups:
hemisphere\,1 $\{ p_{1k} \}$,
hemisphere\,2 $\{ p_{2k} \}$,
which satisfy the following conditions.
\begin{eqnarray}
{\rm Any} \ p_{1i} \in \{ p_{1k} \},
\ p_{2i} \in \{ p_{2k} \}
&&\!\!\!\!\!\!\!
{\rm satisfy\ the\ conditions} 
\cr
\cr
d(p_{1,{\rm ax}},p_{1i})&\le& d(p_{2,{\rm ax}},p_{1i}),\cr
d(p_{2,{\rm ax}},p_{2i})&\le& d(p_{1,{\rm ax}},p_{2i}).
\end{eqnarray}

We define the {\rm ax}sis of hemisphere $p_{{\rm ax}}$ 
and the distance between two 4-vectors $d(p_1,p_2)$ as follows.
\begin{eqnarray}
p_{1,{\rm ax}}\equiv\sum_i p_{1i},\ \ \ p_{2,{\rm ax}} \equiv \sum_i p_{2i},
\label{hemi}
\end{eqnarray}
\begin{eqnarray}
d(p_{{\rm ax}},p_i)&\equiv&
\frac{(E_{{\rm ax}}-|\mathbf{p}_{{\rm ax}}|\cos\theta_{i})E_{{\rm ax}}}{(E_{{\rm ax}} + E_{i})^2}
\ \ \ {\rm (Here,} 
\ \theta_i
\ {\rm is\  the\ angle\ between\ }
\mathbf{p}_{\rm ax}\ {\rm and}\ \mathbf{p}_i).
\end{eqnarray}
Here, high $p_T$ objects mean jets 
with $p_T \ge 50$\,GeV and $\eta \le 3$, 
leptons, photons with 
$p_T\ge 10\,{\rm GeV}$ and $\eta \le 2.5$.

Our algorithm to find the hemisphere axes is as follows.
We take the highest $p_T$ object in the event 
and regard its momentum as $p_{1,{\rm ax}}$.
Next, $p_{2,{\rm ax}}$ is taken as the momentum of the object which has largest 
$|\mathbf{p}|\Delta R$,
where $\mathbf{p}$ is the momentum of an object, 
$\Delta R \equiv \sqrt{(\Delta \eta)^2 + (\Delta \phi)^2} $,
and $\Delta \eta$ and $\Delta \phi$ are the differences of 
the pseudo-rapidity and 
azimuthal
angle 
of $\mathbf{p}$ from $\mathbf{p}_{1,{\rm ax}}$ respectively.
We calculate 
$d(p_{1,{\rm ax}},p)$ and $d(p_{2,{\rm ax}},p)$
for all high $p_T$ objects.
We identify it to hemisphere 1
if $d(p_{1,{\rm ax}},p)<d(p_{2,{\rm ax}},p)$.
Otherwise, to hemisphere 2.
After that,
we redefine $p_{1,{\rm ax}}$ and $p_{2,{\rm ax}}$
as the new hemisphere axes
by using eq.(\ref{hemi}).
We iterate the same operation using new 
$p_{1,{\rm ax}}$ and $p_{2,{\rm ax}}$ five times.

After determination of two hemispheres,
we calculate the maximum invariant mass $m_{jj}$ of all jet pairs in a hemisphere.
We call it $m_{jj1}$ for hemisphere\,1 and $m_{jj2}$ for hemisphere\,2.
When a hemisphere has only one or zero jets,
we define $m_{jj} = 0$.

\begin{figure}[htbp]
\begin{minipage}{5.0cm}
\includegraphics[scale=0.25]{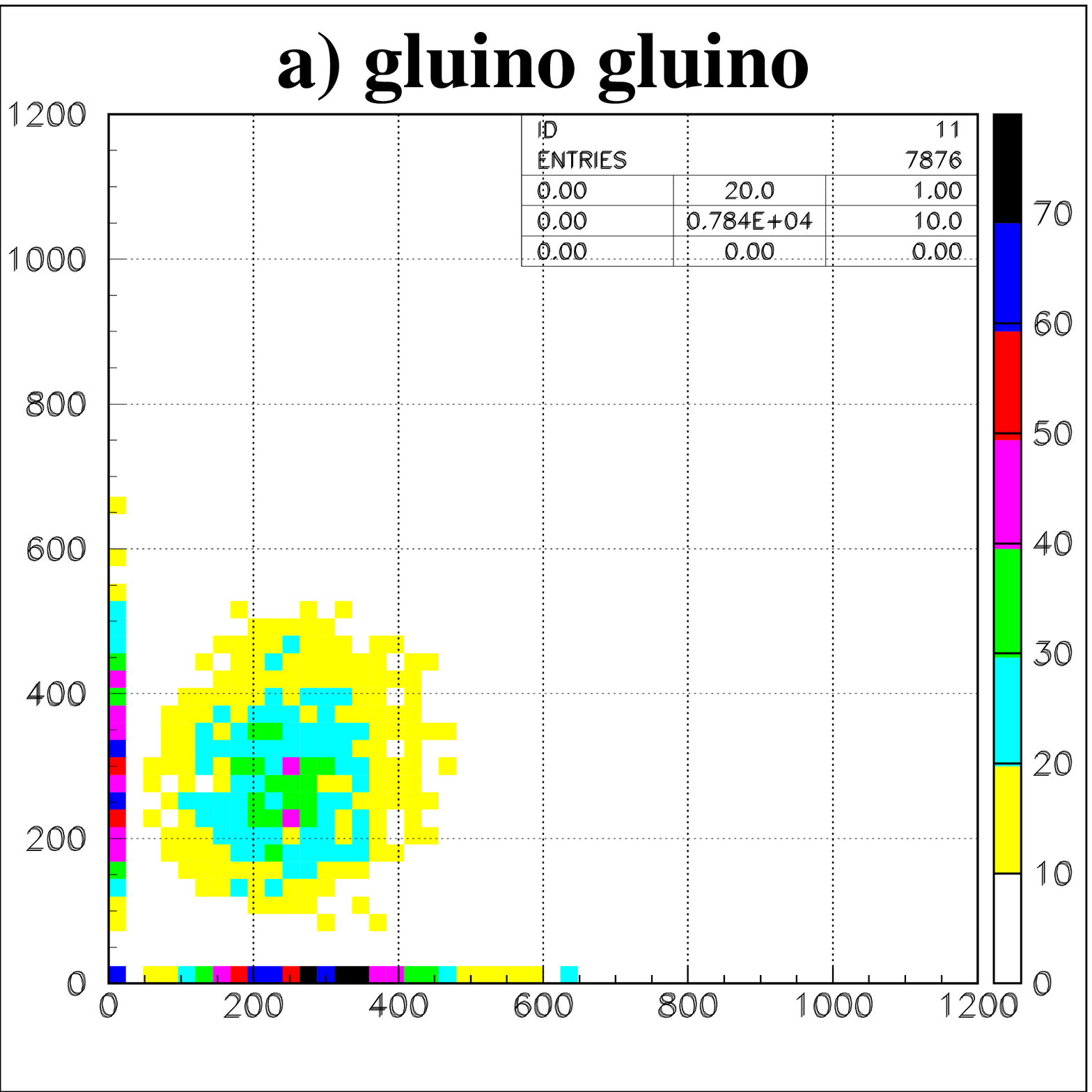}
\end{minipage}
\hfill
\begin{minipage}{5.0cm}
\includegraphics[scale=0.25]{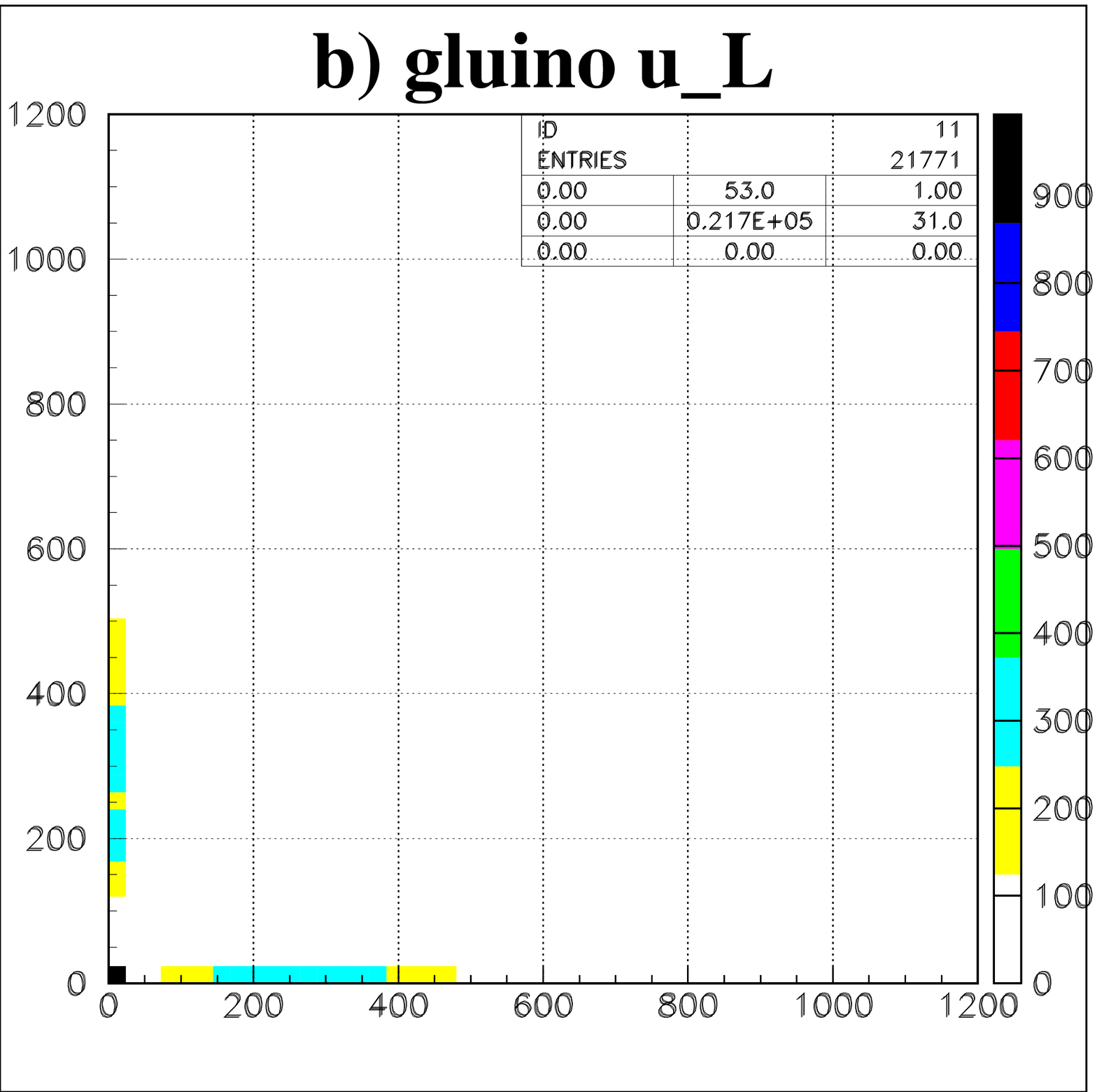}
\end{minipage}
\hfill
\begin{minipage}{5.0cm}
\includegraphics[scale=0.25]{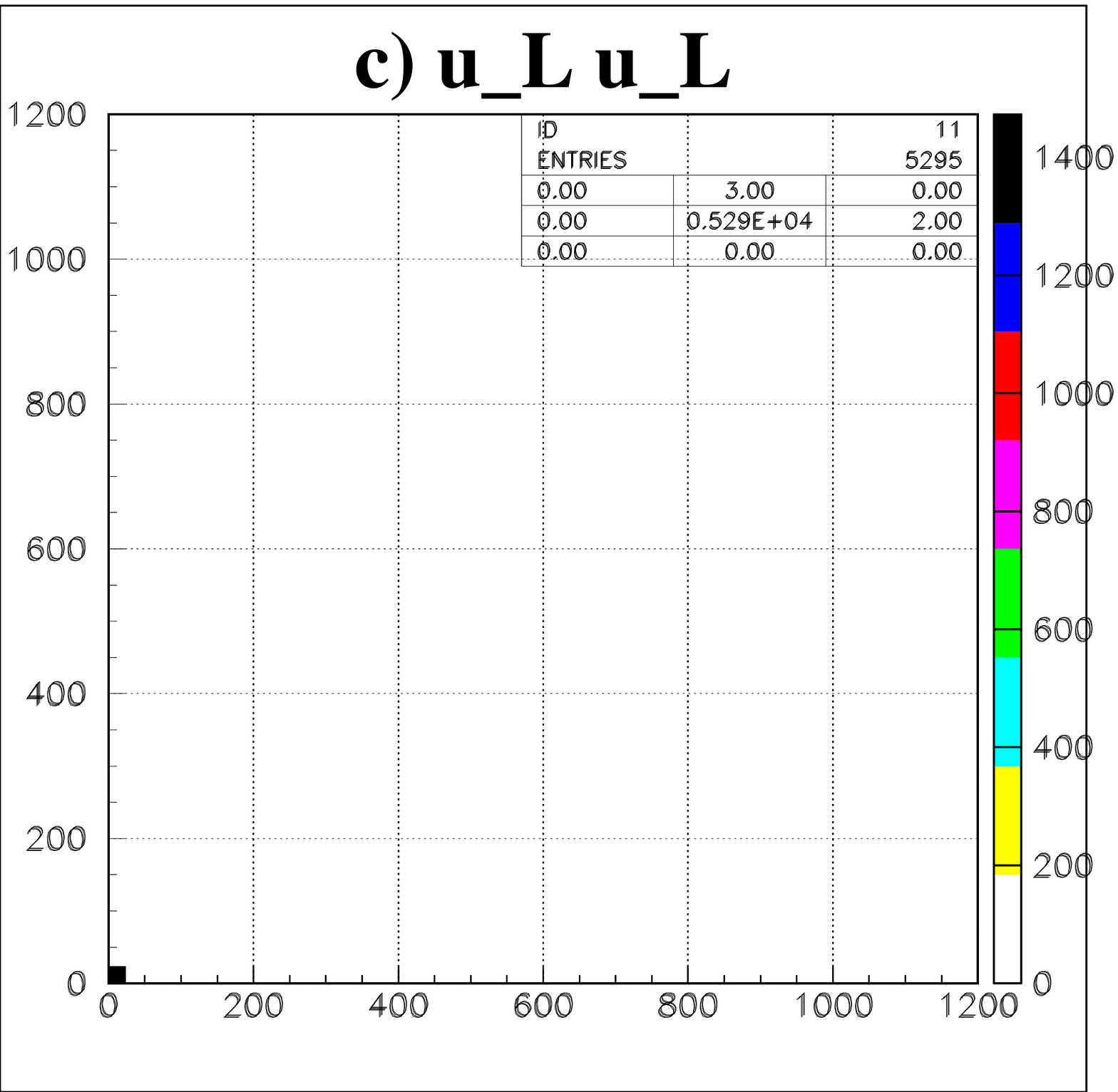}
\end{minipage}
\caption{$m_{jj1}$\,vs.$m_{jj2}$ distribution 
for a) $\gluino\gluino$,
 b) $\gluino\ti u_L$,
 c) $\ti u_L\ti u_L$
 production events.}
\label{mjj}
\end{figure}

The 2-dim plots of $m_{jj1}$ vs. $m_{jj2}$ for production processes 
$\gluino\gluino$, $\gluino\ti u_L$, $\ti u_L\ti u_L$ at Point B
are shown in Figure \ref{mjj}a$\sim$c.
Here, the plotted events are not only SS2$l$ events 
but all events after imposing $c_0$ cut.

We find $\gluino\gluino$ events are mainly distributed in the region of $m_{jj1} \ne 0$
and $m_{jj2} \ne 0$,
$\gluino\sq$ events are mainly distributed in the region of either 
$m_{jj1} = 0$ or $m_{jj2} = 0$,
$\sq\sq$ events are mainly distributed in the region of
 $m_{jj1} = m_{jj2} = 0$.
This is because 
$\gluino$ produces two high $p_T$ jets while $\sq$ produces only one high $p_T$ jet.
Therefore we require $m_{jj1} = m_{jj2} = 0$ as the cut 
to reduce $\gluino$ production events.

\subsection{Numerical results after hemisphere cuts}
We apply the cuts $m_{jji}=0$ and/or a $b$-veto
on the SS2$l$ events and investigate the efficiencies for each production channel.
We subsequently apply the cuts to remove 
the contributions from $\gluino \sq$ production,
while keeping the contributions from $\sq\sq$.
We define cuts $c_2 \sim c_5$ in addition to $c_0$ and $c_1$ as follows,
\begin{eqnarray}
c_2&:& c_0{\rm \ with\  min}(m_{jj1},m_{jj2})=0.\cr
c_3&:& c_0{\rm \ with\  } m_{jj1}=m_{jj2}=0.\cr
c_4&:& c_0,n_b=0{\rm \ and\  min}(m_{jj1},m_{jj2})=0.\cr
c_5&:& c_0,n_b=0{\rm \ and\ } m_{jj1}=m_{jj2}=0. \cr
{\rm ratio}&:& N(c_5)/N(c_0).
{\rm\ This\ describes\ the\ efficiency\ for}\ c_5 \ {\rm cut.}\nonumber
\end{eqnarray}
The number of the SS2$l$ events after these cuts 
are shown in the Table \ref{spsmis500}.
The cut requiring $m_{jj1}=0$ or $m_{jj2}=0$ ($c_2,c_4$) reduces 
$\gluino\gluino$ events drastically and $\gluino \sq$ events moderately.
The cut requiring $m_{jj1}=0$ and $m_{jj2}=0$ ($c_3,c_5$) 
further reduces $\gluino\sq$ events.
On the other hand, $\sq\sq$ events survive under the cut $c_5$ 
compared with $\gluino\gluino,\ \gluino \sq$ events.
Note that a $b$-veto cut is not essential to reduce the gluino contribution.
The cut $c_3$ reduces events involving gluino less than $\sq_L\sq_L$ events
although the $b$-veto is not applied.
$\gluino \sq$ events are dominant under the $c_0$ cut and 
$\sq \sq$ events are dominant under the $c_5$ cut.

\begin{table}[htbp]
\small
\center{
\begin{tabular}{|cc||r|rrrrrr|r|}
\hline 
\multicolumn{2}{|c||}{Point A} & all & $c_0$ & $c_1$  & $c_2$                                                           & $c_3$ & $c_4$ & $c_5$ & ratio \cr
\hline 
$l^+ l^+ $& total        &  1519&   716&   349&   463&   114&   267&    94& 0.131\cr
  & $\sq\sq$             &   235&   140&   132&   123&    57&   117&    56& 0.400\cr
  & $\gluino \sq$        &   618&   333&   125&   205&    27&    83&    16& 0.048\cr
  & $\gluino\gluino$     &   184&    79&    14&    27&     2&     5&     0& 0.000\cr
\hline 
$l^- l^- $& total        &  1213&   610&   286&   368&    89&   216&    75& 0.123\cr
  & $\sq\sq$             &   151&    92&    90&    83&    42&    81&    41& 0.446\cr
  & $\gluino \sq$        &   472&   262&   108&   159&    19&    75&    13& 0.050\cr
  & $\gluino\gluino$     &   172&    93&    20&    31&     1&     6&     0& 0.000\cr
\hline
\multicolumn{10}{c}{ }\cr
\hline 
\multicolumn{2}{|c||}{Point B} & all & $c_0$ & $c_1$  & $c_2$                                                           & $c_3$ & $c_4$ & $c_5$ & ratio \cr
\hline 
$l^+ l^+ $& total        &  4363&  2573&  1612&  1765&   465&  1293&   441& 0.171\cr
  & $\sq\sq$             &  1479&  1014&  1001&   894&   363&   883&   361& 0.356\cr
  & $\gluino \sq$        &  1765&  1098&   433&   613&    52&   279&    36& 0.033\cr
  & $\gluino\gluino$     &   220&   117&    18&    26&     1&     5&     0& 0.000\cr
\hline 
$l^- l^- $& total        &  2231&  1288&   708&   809&   187&   561&   171& 0.133\cr
  & $\sq\sq$             &   499&   326&   316&   286&   122&   278&   121& 0.371\cr
  & $\gluino \sq$        &   861&   541&   237&   304&    23&   169&    15& 0.028\cr
  & $\gluino\gluino$     &   225&   121&    22&    32&     0&    11&     0& 0.000\cr
\hline
\end{tabular}                     
}                                 
\end{table}                       
\begin{table}[htbp]               
\center{                          
\begin{tabular}{|cc||r|rrrrrr|r|} 
\hline 
\multicolumn{2}{|c||}{Point C} & all & $c_0$ & $c_1$  & $c_2$                                                           & $c_3$ & $c_4$ & $c_5$ & ratio \cr
\hline 
$l^+ l^+ $& total        &  1081&   467&   242&   284&    66&   186&    62& 0.133\cr
  & $\sq\sq$             &   259&   154&   144&   133&    57&   127&    57& 0.370\cr
  & $\gluino \sq$        &   327&   165&    55&    69&     3&    29&     2& 0.012\cr
  & $\gluino\gluino$     &    47&    10&     1&     2&     0&     0&     0& 0.000\cr
\hline 
$l^- l^- $& total        &   618&   233&   110&   134&    20&    83&    17& 0.073\cr
  & $\sq\sq$             &    77&    40&    39&    37&    12&    36&    12& 0.300\cr
  & $\gluino \sq$        &   164&    66&    23&    29&     2&    14&     0& 0.000\cr
  & $\gluino\gluino$     &    32&    11&     3&     6&     0&     2&     0& 0.000\cr
\hline
\multicolumn{10}{c}{ }\cr
\hline 
\multicolumn{2}{|c||}{Point D} & all & $c_0$ & $c_1$  & $c_2$                                                           & $c_3$ & $c_4$ & $c_5$ & ratio \cr
\hline 
$l^+ l^+ $& total        &  1157&   571&   318&   335&    82&   243&    78& 0.137\cr
  & $\sq\sq$             &   356&   216&   207&   182&    72&   179&    72& 0.333\cr
  & $\gluino \sq$        &   333&   169&    56&    64&     2&    26&     2& 0.012\cr
  & $\gluino\gluino$     &    35&    18&     5&     7&     0&     4&     0& 0.000\cr
\hline 
$l^- l^- $& total        &   588&   263&   116&   128&    27&    83&    24& 0.091\cr
  & $\sq\sq$             &    84&    55&    47&    45&    16&    40&    16& 0.291\cr
  & $\gluino \sq$        &   131&    63&    25&    23&     1&    14&     1& 0.016\cr
  & $\gluino\gluino$     &    27&     9&     0&     0&     0&     0&     0& 0.000\cr
\hline
\multicolumn{10}{c}{ }\cr
\hline 
\multicolumn{2}{|c||}{sps1a}  & all & $c_0$ & $c_1$  & $c_2$                                                           & $c_3$ & $c_4$ & $c_5$ & ratio \cr
\hline 
$l^+ l^+ $& total        &  1314&   461&   250&   326&    80&   193&    63& 0.137\cr
  & $\sq\sq$             &   164&    72&    71&    59&    29&    58&    28& 0.389\cr
  & $\gluino \sq$        &   550&   226&   106&   150&    22&    75&    15& 0.066\cr
  & $\gluino\gluino$     &   155&    58&    15&    32&     2&     9&     2& 0.034\cr
\hline 
$l^- l^- $& total        &  1070&   345&   180&   242&    64&   147&    46& 0.133\cr
  & $\sq\sq$             &   109&    47&    45&    41&    16&    40&    16& 0.340\cr
  & $\gluino \sq$        &   394&   156&    64&   109&    27&    51&    16& 0.103\cr
  & $\gluino\gluino$     &   147&    39&     9&    17&     2&     7&     0& 0.000\cr
\hline
\end{tabular}
\caption{Numbers of SS2$l$ events after the cuts for Point A$\sim$D and SPS1a.
Here, $c_0$ is set to $E\slush_T> 200\,$GeV, $M_{\rm eff} > 500\,$GeV, 
$E\slush_T> 0.2 M_{\rm eff}$, $n_{100} \ge 2$.}
\label{spsmis500}
}
\end{table}

The ratio of $N(l^+l^+;{c_5})$/$N(l^+l^+;{c_0})$ for $\sq\sq$ productions is more than 30\%
at our model points.
It is $3\sim 5$\% for $\gluino\sq$ production at Point A and B 
and about 1\% at Point C and D.
We can obtain a pure SS2$l$ event set from $\sq\sq$ production by the $c_5$ cut.
Dominant contributions to $N(l^+l^+$ from $\sq\sq)$ 
are from $\sq_L\sq_L$ production (mainly $\ti u_L\ti u_L$).
We can see in Appendix \ref{dominantcontribution},
$N(l^+l^+$ from $\ti u_L\ti u_L$ and $\ti u_L\ti c_L$)=1256,
$N(l^+l^+$ from $\ti u_L\ti d_L^\ast$ and $\ti u_L\ti s_L^\ast$)=138,
$N(l^+l^+$ from $\ti u_L\ti d_L$)=84
are obtained among $N(l^+l^+$ from all $\sq\sq$)=1479 at Point B.
In Table \ref{spsmis500}, we also show the results at SPS1a for a reference.
The efficiencies of these cuts are similar to the other points.
However contamination from $\gluino \sq$ production is larger.

Experimentally, we can only observe the total number of SS2$l$ events.
The ratio $N(l^+l^+;{c_5})/N(l^+l^+;{c_0})$ 
for total events is 13$\sim$17\% at our points. 
If there is no $\sq_L\sq_L$ production,
it becomes less than 5\%.
If the efficiency of $N(l^+l^+;{c_5})/N(l^+l^+;{c_0})$ for $\sq\sq$ and $\gluino\sq$
can be obtained by MC simulations and from the other constraints,
we may estimate $\sigma(\sq_L\sq_L)$ from $N(l^\pm l^\pm)$.
The model parameters we need are the gluino, squark masses 
to constrain the branching ratio of the gluino,
leptonic branching ratios of squarks and their decay kinematics.
The ratio of events with 2 leptons to events with 1 lepton
should be useful to estimate the leptonic branching ratio of $\sq$.
The decay cascade $\gluino \to \ti b \bar{b} (\ti t \bar{t}) \to \ch^- t\bar{b}(\ch^+ \bar{t}b)$
also emits leptons from $W^\pm$ decays.
The branching ratio must be estimated carefully from $b$-tagged samples.
Estimation of the errors on the branching ratio is beyond the scope of this paper.
Information on the mass would be obtained 
from various end points of the decay distributions 
\cite{Hinchliffe:1996iu,Bachacou:1999zb,Hinchliffe:1999zc,Hinchliffe:1998ys,Hisano:2003qu,Kawagoe:2004rz}.
In addition, there are various sources of systematic errors 
for the cross sections 
such as the PDF, NNLO corrections and so on.
Following the arguments in \cite{Freitas:2006wd}, we assume that the overall uncertainty 
on the total production cross section is 17\%,
which comes from 10\% PDF uncertainty, 8\% NNLO corrections \cite{NNLO} and
10\% $\sigma(\sq_L \sq_L)$ error from a 3\% squark mass uncertainty.
If the squark mass error is 10\%, $\sigma(\sq_L \sq_L)$ error is 40\%, 
and this uncertainty dominates the systematic errors.

Uncertainties on the cross sections may partly cancel in 
$N(l^\pm l^\pm;{c_5})/N(l^\pm l^\pm;{c_0})$ ratios.
The errors on the absolute sparticle masses are 
common for both $\sigma(\sq_L\sq_L)$ and $\sigma(\gluino\sq_L)$,
and $m_{\gluino} - m_{\sq}$ may be known precisely.
If the cross section errors cancel,
the total systematic error becomes 14\%.
Uncertainties on leptonic branching ratios of squarks
may also partly cancel 
because the gluino emits leptons through its decay into a squark.

Note that the ratio of the production cross sections of
$\gluino \ti u_L$ to $\gluino \ti d_L$ is 2:1,
and that of $\ti u_L \ti u_L$ to $\ti d_L \ti d_L$ is 4:1.
Indeed, at Points C and D,
the ratio $N(l^+l^+;c_5)$ to $N(l^-l^-;c_5)$
is 4:1, while that is 2:1 for $c_0$. 
At Point A (B),
the ratio of $N(l^+l^+;c_5)$ to $N(l^-l^-;c_5)$ is 5:4 (5:2),
while the ratio of $N(l^+l^+;c_0)$ to $N(l^-l^-;c_0)$ is 7:6 (2:1).
These discrepancies from $\sigma(\sq^+\sq^+)$/$\sigma(\sq^-\sq^-)$ 
or  $\sigma(\gluino\sq^+)$/$\sigma(\gluino\sq^-)$ in Table \ref{gene} are caused 
by decays of the second lightest neutralinos from $\sq_L^\pm$.
According to Appendix \ref{appbranch}, 
the left handed squark can decay into $\none \tau^+\tau^-$ through $\ntwo$
with a large branching ratio at Point A (B).
Therefore there is a contamination from 
$\sq_L^+ \sq_L^-$ (mainly $\ti u_L \ti d_L$) production to SS2$l$ events
and the ratio of $N(l^+l^+)$ to $N(l^-l^-)$ gets closer to 1:1.
Therefore the charge of the hard lepton does not reflect 
the sign of the parent particle.

\subsection{Tight cut for heavy mass spectrum}
We now consider the SM background to SS2$l$ events.
Here we only consider $t\bar{t}$ production
which is found as dominant background in \cite{Freitas:2006wd}.
The background comes from the events where 
one lepton comes from the leptonic decay of a top quark
while the other lepton comes from accidental sources such as 
$b$ quark decays.
The ${\cal BR}(t \to l^-)$ is small, however the total number of $t\bar{t}$ events
is significantly larger than the signal 
($\sigma(t\bar{t}) = 400\,$pb at tree level). 
The $t\bar{t}$ events corresponding to $100\,$fb$^{-1}$ are also generated
by HERWIG\,6.5,
and the result is shown in Table \ref{tt1}.
\begin{table}[htbp]
\center{
\small
\begin{tabular}{|c||r|rrrrrr|r|}
\hline 
$ t\bar{t}$ & all & $c_0$ & $c_1$  & $c_2$                                                           & $c_3$ & $c_4$ & $c_5$ & ratio \cr
\hline 
$l^+ l^+$                &  1710&   152&    70&   133&    32&    65&    17& 0.112\cr
\hline 
$l^- l^-$                &  1635&   146&    76&   133&    20&    70&    10& 0.068\cr
\hline
\end{tabular}
\caption{Numbers of SS2$l$ events from $t\bar{t}$ production 
after the cuts $c_0 \sim c_5$ for $\int\!\!dt {\cal L} = 100$fb$^{-1}$.
Here, $c_0 \sim c_5$ are the same as Table 6.}
\label{tt1}}
\end{table}

At Points A and B,
$N(l^+l^+$ from $t\bar{t};c_0)$ is less than 8\% 
of $N(l^+l^+$ from SUSY;$c_0)$
and $N(l^+l^+$ from $t\bar{t};c_5)$ is about 5\% of $N(l^+l^+$ from SUSY;$c_5)$.
At Point C,
it is 30\% after $c_0$,
and 25\% after $c_5$.
At Point D, 
they are $\sim 1$.
In that case, more strict $E\slush_T$ and $M_{\rm eff}$ cuts are needed.

In Table \ref{spsmis500}, we have taken $E\slush_{T} > 200$\,GeV, $M_{\rm eff} > 500$\,GeV
and $E\slush_{T} > 0.2M_{\rm eff}$ as the $c_0$ cut.
Numbers of $t\bar{t}$ after various 
$E\slush_T$ and $M_{\rm eff}$ cuts are shown in Table \ref{ttback} .
The $t\bar{t}$ background is dramatically reduced 
if high $E\slush_T$ cuts are applied for large $M_{\rm eff}$ \cite{hunting}.
Here,
we change the $c_0$ cut as follows.
\\
\mbox{\boldmath $E\slush_T$} {\bf and} \mbox{\boldmath $M_{\rm eff}$} {\bf cuts}
\begin{eqnarray}
c^{(1)}&:&E\slush_T > 200 \,{\rm GeV}
,\ E\slush_T > 0.2 M_{\rm eff},\ M_{\rm eff} > 500 \,{\rm GeV}\cr
c^{(2)}&:&E\slush_T > 250 \,{\rm GeV}
,\ E\slush_T > 0.2 M_{\rm eff},\ M_{\rm eff} > 750 \,{\rm GeV}\cr
c^{(3)}&:&E\slush_T > 300 \,{\rm GeV}
,\ E\slush_T > 0.2 M_{\rm eff},\ M_{\rm eff} > 1000 \,{\rm GeV}\nonumber
\end{eqnarray}
{\bf Number of high $p_T$ jets}
\begin{eqnarray}
n_{100} \ge 2&:&{\rm at\  least\  two\  jets\  with\ } p_T > 100 \,{\rm GeV}\cr
n_{200} \ge 2&:&{\rm at\  least\  two\  jets\  with\ } p_T > 200 \,{\rm GeV}\nonumber
\end{eqnarray}
\begin{table}[htbp]
\center{
\small
\begin{tabular}{|c||r|rrr|rrr|}
\hline 
number of jets & all  &\multicolumn{3}{c|}{$n_{100} \ge 2 $ } & \multicolumn{3}{c|}{$n_{200} \ge 2 $ } \cr
\hline 
$E\slush_T ,M_{\rm eff}$ cut & all & $c^{(1)}$ &  $c^{(2)}$ & $c^{(3)}$ & $c^{(1)}$ &  $c^{(2)}$ & $c^{(3)}$\cr
\hline
\hline 
$N(l^+ l^+ {\rm \ from} \  t\bar{t} ;c_0)$  &  1710 &  152 &  38 &  9 & 13 &   6 &  4  \cr
$N(l^+ l^+ {\rm \ from} \  t\bar{t} ;c_5)$  &     * &   17 &   1 &  0 &  3 &   0 &  0  \cr
\hline 
$N(l^- l^- {\rm \ from} \  t\bar{t} ;c_0)$    &  1635 &  146 & 43 &  15 &  11 &   8 & 2  \cr
$N(l^- l^- {\rm \ from} \  t\bar{t} ;c_5)$    &     * &   10 &  2 &   1 &   0 &   0 & 0  \cr
\hline
\end{tabular}
\caption{Number of SS2$l$ events from $t\bar{t}$ events 
after various $c_0$ cuts for 100\,fb$^{-1}$.}
\label{ttback}
}
\end{table}

We also show $N(l^\pm l^\pm;{c_5})/N(l^\pm l^\pm;{c_0})$
after the cut $c^{(i)}$ and the cuts on the number of high $p_T$ jets 
at Point D in Table \ref{ttBp}.
The numbers of events correspond to 307.06\,pb$^{-1}$.
We can see that the efficiency $N(l^\pm l^\pm;{c_5})/N(l^\pm l^\pm;{c_0})$ 
for $\sq\sq$ production
and the efficiency for $\gluino\sq$ production
only weakly depend on the basic cut 
$c^{(i)}$ and the cuts on the number of high $p_T$ jets.
The $c_5$ cut is still useful to reduce $\gluino \sq$ productions.

We can drop the $t\bar{t}$ background without reducing the signal from SUSY events
so much by taking $c^{(3)}$ and $n_{200} \ge 2$ as $c_0$.
However, it has recently been pointed out that the number of high $p_T$ jets
increases significantly if matrix element (ME) corrections are included \cite{asai}.
On the other hand, the $E\slush_T$ cut is not affected by ME corrections.
The $c^{(3)}$ cut reduces background efficiently.

\begin{table}[htbp]
\center{
\small
\begin{tabular}{|c||r|rrr|rrr|}
\hline 
number of jets & all  &\multicolumn{3}{c|}{$n_{100} \ge 2 $ } & \multicolumn{3}{c|}{$n_{200} \ge 2 $ } \cr
\hline 
$E\slush_T ,M_{eff}$ cut & all & $c^{(1)}$ &  $c^{(2)}$ & $c^{(3)}$ & $c^{(1)}$ &  $c^{(2)}$ & $c^{(3)}$\cr
\hline
\hline 
$N(l^+ l^+ {\rm\ from\ }\sq\sq;c_0)$    &  356 & 216  & 211 & 197 & 162 & 159 & 150 \cr
$N(l^+ l^+ {\rm\ from\ }\gluino\sq;c_0)$&  333 & 169  & 169 & 161 & 122 & 122 & 119 \cr
\hline
$N(l^+ l^+ {\rm\ from\ }\sq\sq;c_5)$    & * & 72  & 67 & 61 & 59 & 56 & 52 \cr
$N(l^+ l^+ {\rm\ from\ }\gluino\sq;c_5)$ & * &  2  &  2 &  2 &  2 &  2 &  2 \cr
\hline
$N(l^+ l^+;{c_5})/N(l^+ l^+;{c_0}) {\rm \ for\ } \sq\sq$ & *& 0.333 & 0.318 & 0.310 & 0.364 & 0.352 & 0.347 \cr
$N(l^+ l^+;{c_5})/N(l^+ l^+;{c_0}) {\rm \ for\ } \gluino\sq$ & *& 0.012 & 0.012 & 0.012 & 0.016 & 0.016 & 0.017 \cr
\hline 
\hline 
$N(l^- l^- {\rm\ from\ }\sq\sq;c_0)$ & 84 & 55  & 54  & 50  & 38 & 37 & 36  \cr
$N(l^- l^- {\rm\ from\ }\gluino\sq;c_0)$ & 131& 63  &60   &56   & 47 & 47 & 46\cr
\hline
$N(l^- l^- {\rm\ from\ }\sq\sq;c_5)$       & * & 16  & 15  & 15 & 14& 13 &13 \cr
$N(l^- l^- {\rm\ from\ }\gluino\sq;c_5)$ & * &  1 &  1 & 1 &1 & 1 &1 \cr
\hline
$N(l^- l^-;{c_5})/N(l^- l^-;{c_0}) {\rm \ for\ } \sq\sq$   & *&0.291&0.278 & 0.300  &0.368 &0.351 & 0.361 \cr
$N(l^- l^-;{c_5})/N(l^- l^-;{c_0}) {\rm \ for\ } \gluino\sq$& *& 0.016  & 0.017  & 0.018 & 0.021& 0.021& 0.022\cr
\hline
\end{tabular}
\caption{Number of SS2$l$ events at Point D after various $c_0$ cuts for 307.06 fb$^{-1}$}
\label{ttBp}
}
\end{table}

\subsection{Summary of the cuts}

It is important to reduce the background from $\gluino\sq$ production 
to measure the $\sq_L\sq_L$ production cross section using SS2$l$ events.
We give a systematic procedure to separate $\gluino$ and $\sq$
based on the number of jets in a hemisphere,
and demonstrate that it works well 
to separate $\sq_L\sq_L$ from $\gluino\sq$ 
for our model points.
The hemisphere cut should work provided that $m_{\gluino} - m_{\sq}$
is sufficiently large that a jet from the decay $\gluino \to \sq q$
is detectable.

Using the cut on the number of jets in a hemisphere and the $b$-jet veto,
SS2$l$ events from $\gluino\sq$ production are reduced by more than 95\%
while SS2$l$ events from $\sq_L\sq_L$ are 
selected with an efficiency of
more than 30\%
at our model points.
Moreover, these efficiencies depend only weakly on the basic cuts
on $E\slush_T$ and $M_{\rm eff}$.

Evidence of squark pair production can be seen 
in the ratio of the events
before and after the hemisphere cuts,
because it is significantly different from 
that of $\gluino\sq$ production.
At Point B,
the ratio $N(l^+l^+;c_5)$/$N(l^+l^+;c_0)$
is 0.171 with a statistical error of $\sim$0.01.
If $\sq\sq$ production does not occur,
$N(l^+l^+;c_0)$ becomes 1559,
 $N(l^+l^+;c_5)$ becomes 80
and the ratio becomes 0.051 with a statistical error of $\sim$0.006.
Actually, there are various sources of systematic errors such as
uncertainties on the PDF, NNLO corrections and so on.
The total systematic error is 14\%
if the uncertainties from squark and gluino mass errors 
cancel in the ratio
and we ignore errors on the branching ratios.
However we think 
there is enough margin to identify $\sq_L\sq_L$ production
if careful analyses are done at LHC.

\section{Comparison of other models with the MSSM}
\subsection{The model with an extended gluino sector}

\begin{figure}[htb]
\includegraphics[scale=0.6]{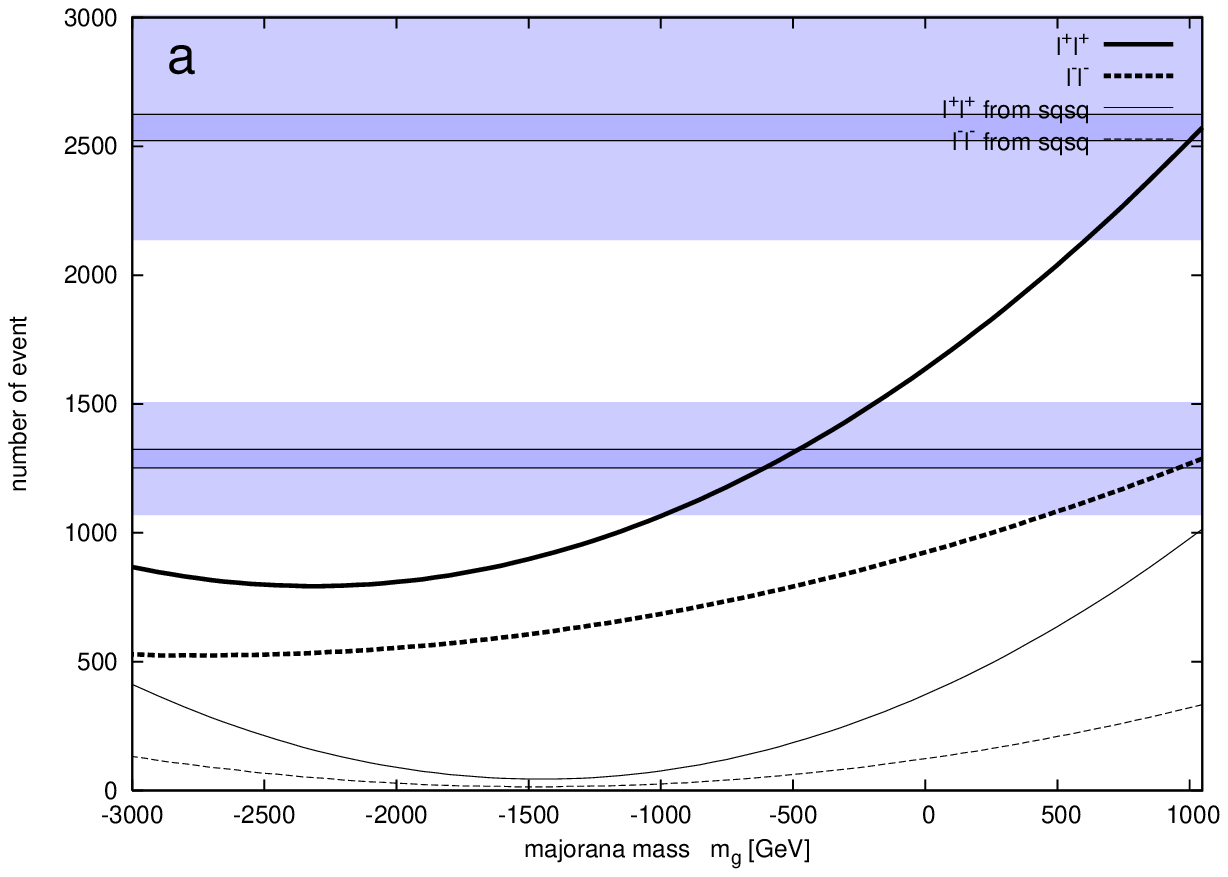}
\hfill
\includegraphics[scale=0.6]{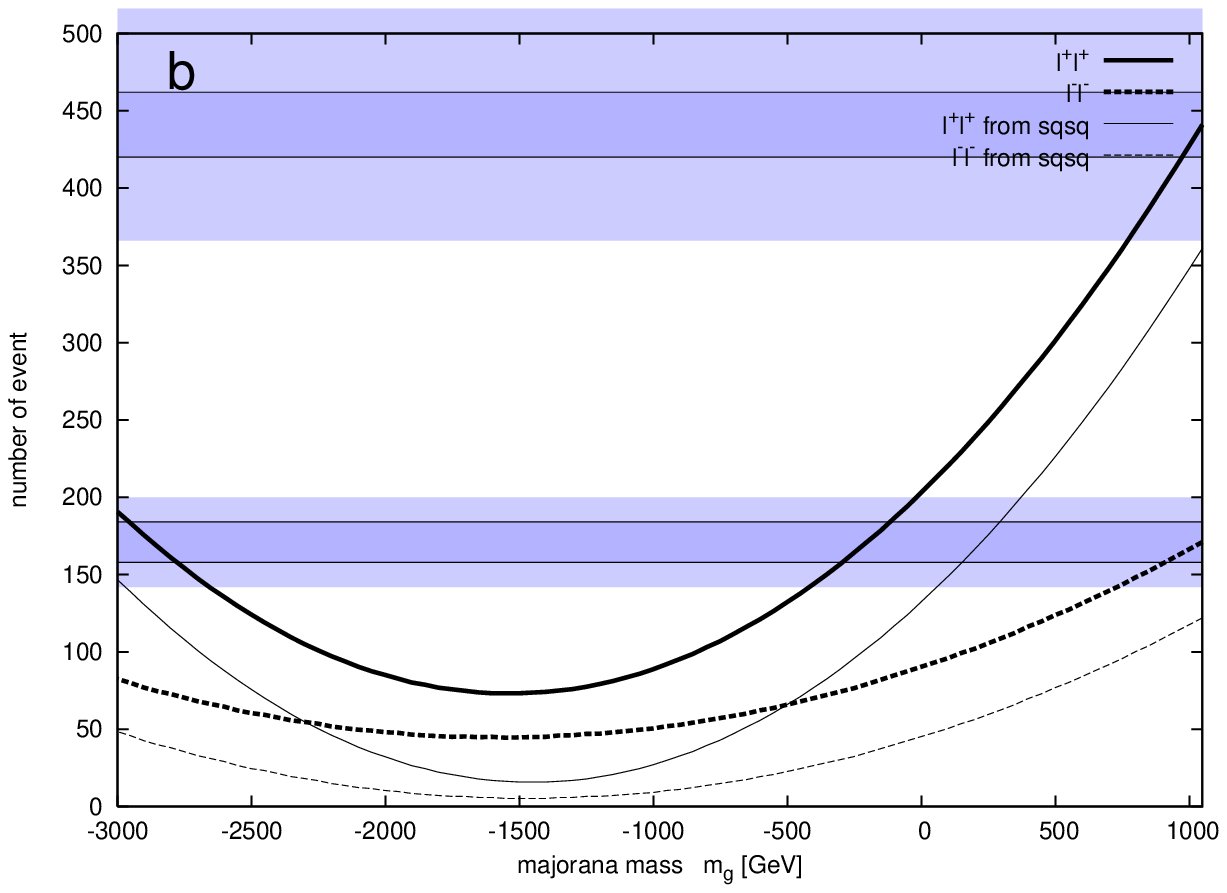}
\caption{Number of SS2$l$ events as a function of the gluino
majorana mass at Point B
a) after the cut $c_0$,
b) after the cut $c_5$.
Dark gray zones show 1$\sigma$ statistical errors for 3$\times 10^5$ events.
Light gray zones show 17\% errors.}
\label{Bhikaku}
\end{figure}

The model with an extended gluino sector 
has been discussed in section 2.2.
In this model, a gluino acquires a Dirac mass term with an adjoint fermion $\ti a$.
As the majorana gluino mass parameter decreases from the MSSM value
for the same gluino mass,
the total SUSY production cross section decreases.
In particlular, $\sigma(\sq_L\sq_L)$ decreases more rapidly than 
$\sigma(\gluino \sq_L)$ when the majorana gluino mass parameter is reduced.
Then SS2$l$ events from $\sq_L\sq_L$ decrease more than
those from $\gluino \sq_L$.
Figure \ref{Bhikaku}a shows $N(l^\pm l^\pm;c_0)$ 
as a function of the majorana gluino mass.
Here, we set the mass spectrum of this model
as that of Point B and $m_{\gluino_1}=m_{\gluino}$ 
and $m_{\gluino_2} = -3000\,$GeV,
and branching ratios and efficiencies of cuts 
are the same as in the previous section.
Moreover, we simplify our calculation by assuming that 
all $l^\pm l^\pm$ events from $\sq\sq$ production 
occur from $\sq^+_L \sq^+_L$ productions 
(see Appendix \ref{dominantcontribution})
and that the non-$\gluino$, non-$\sq_L$ contribution does not depend on $m_g$.
We show $N(l^+l^+;c_0)$ as a bold solid line,
$N(l^-l^-;c_0)$ as a bold dashed line,
$N(l^+l^+$ from $\sq\sq;c_0)$ as a thin solid line
and $N(l^-l^-$ from $\sq\sq;c_0)$ as a thin dashed line.

Figure \ref{Bhikaku}b shows $N(l^\pm l^\pm ; c_5)$.
SS2$l$ events from $\sq_L^+\sq_L^+$ are dominant in the total SS2$l$ events
after the $c_5$ cut,
and that the dependency on the majorana gluino mass of the total number of SS2$l$ events
is nearly the same as the $\sigma(\sq_L\sq_L)$ dependence shown in Figure \ref{diracB}.
On the other hand,
the $\gluino\sq$ contribution is dominant in Figure \ref{Bhikaku}a.

We show the $\pm 1\sigma$ statistical error for the MSSM limits with a dark gray zone.
The absolute numbers of SS2$l$ events after the cut $c_0, c_5$
depend on various parameters, acceptance and so on.
We assume the total uncertainty is 17\% as discussed in the previous section.
This is also shown in Figure \ref{Bhikaku} with a light gray zone.

The $N(l^+l^+ (l^-l^-);c_0)$ shows more than 17\% deviation from the MSSM
in case of $m_g \le 606\,$GeV ($m_g \le 459\,$GeV).
This means $m_D \ge 1262\,$GeV ($m_D \ge 1427\,$GeV). 
It is $m_g \le 936\,$GeV ($m_g \le 730\,$GeV) for $N(l^+l^+ (l^-l^-);c_5)$. 
This means $m_D \ge 664\,$GeV ($m_D \ge 1089\,$GeV). 

\begin{figure}[htb]
\begin{minipage}{8.0cm}
\includegraphics[scale=0.6]{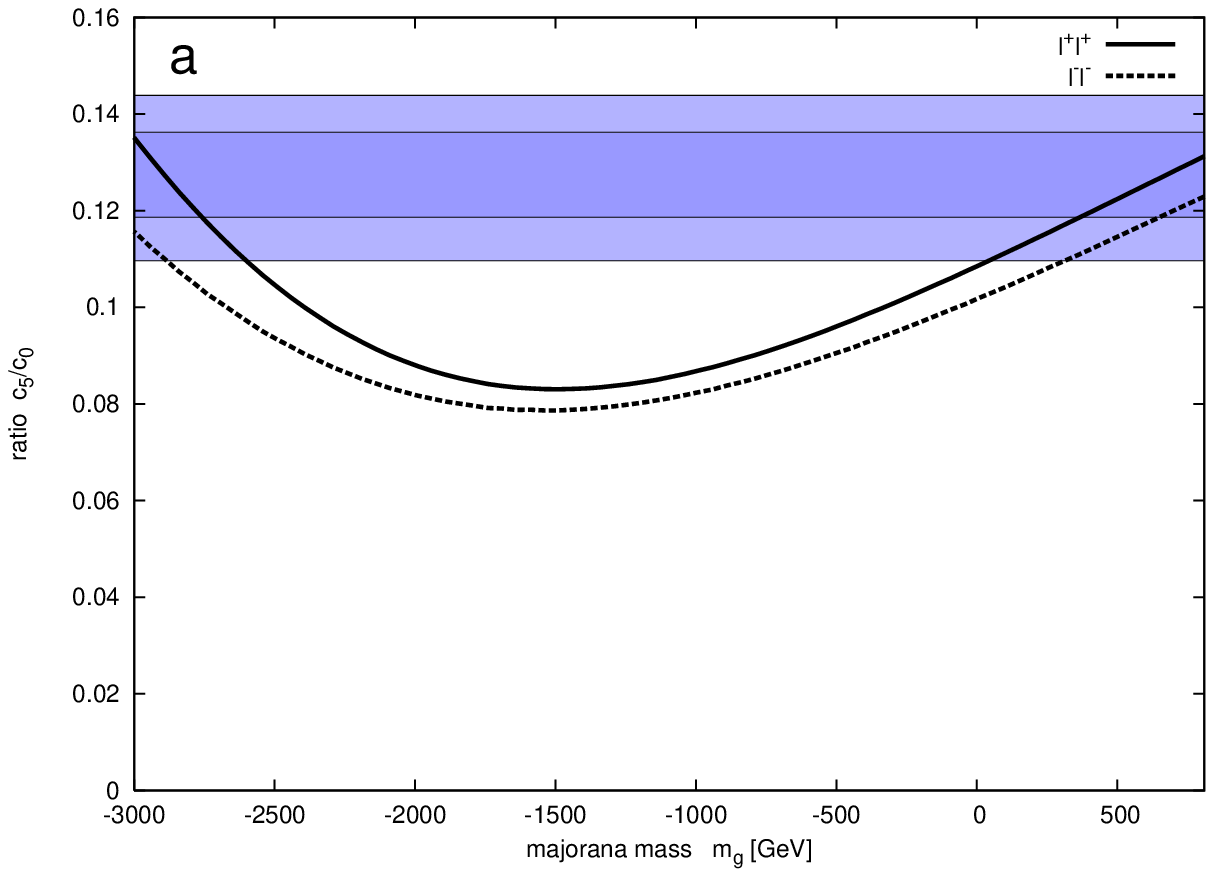}
\end{minipage}
\hfill
\begin{minipage}{8.0cm}
\includegraphics[scale=0.6]{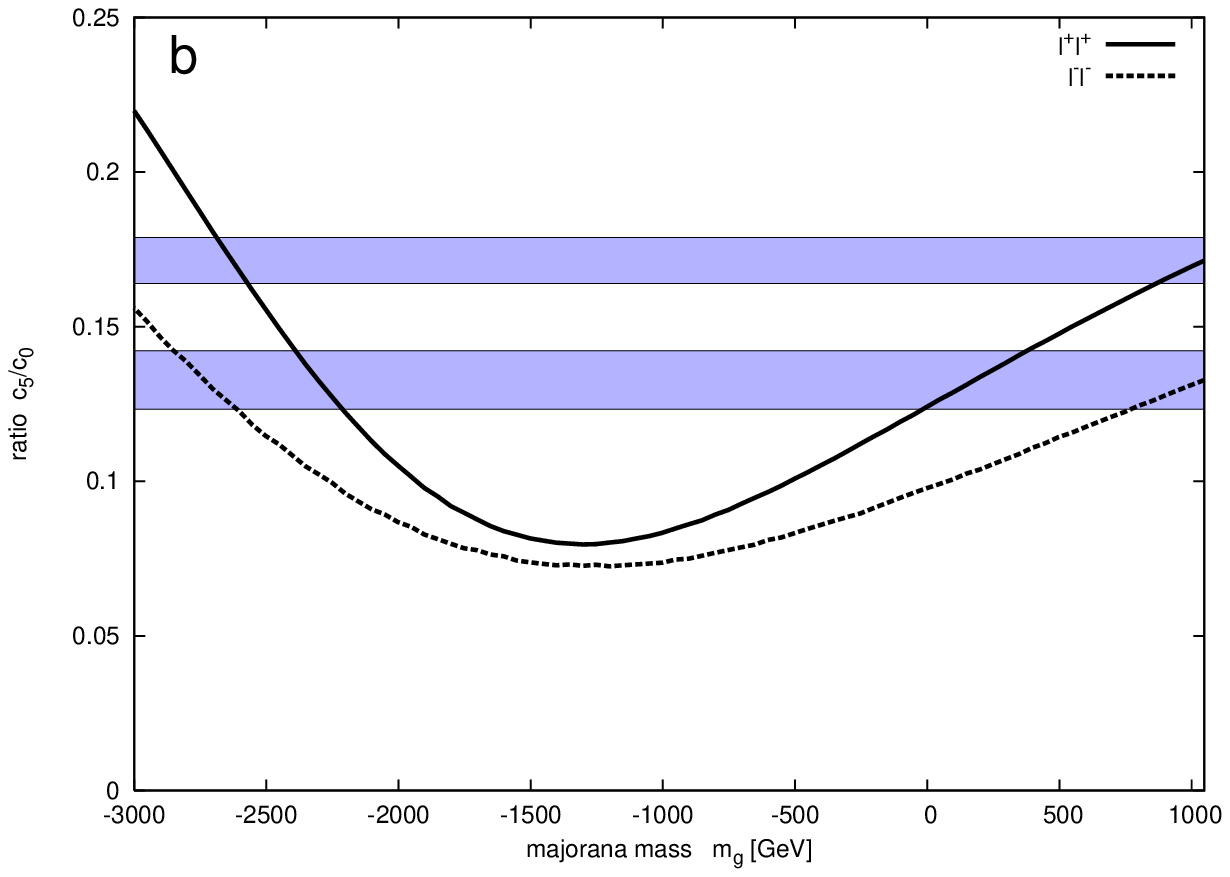}
\end{minipage}
\\
\begin{minipage}{8.0cm}
\includegraphics[scale=0.6]{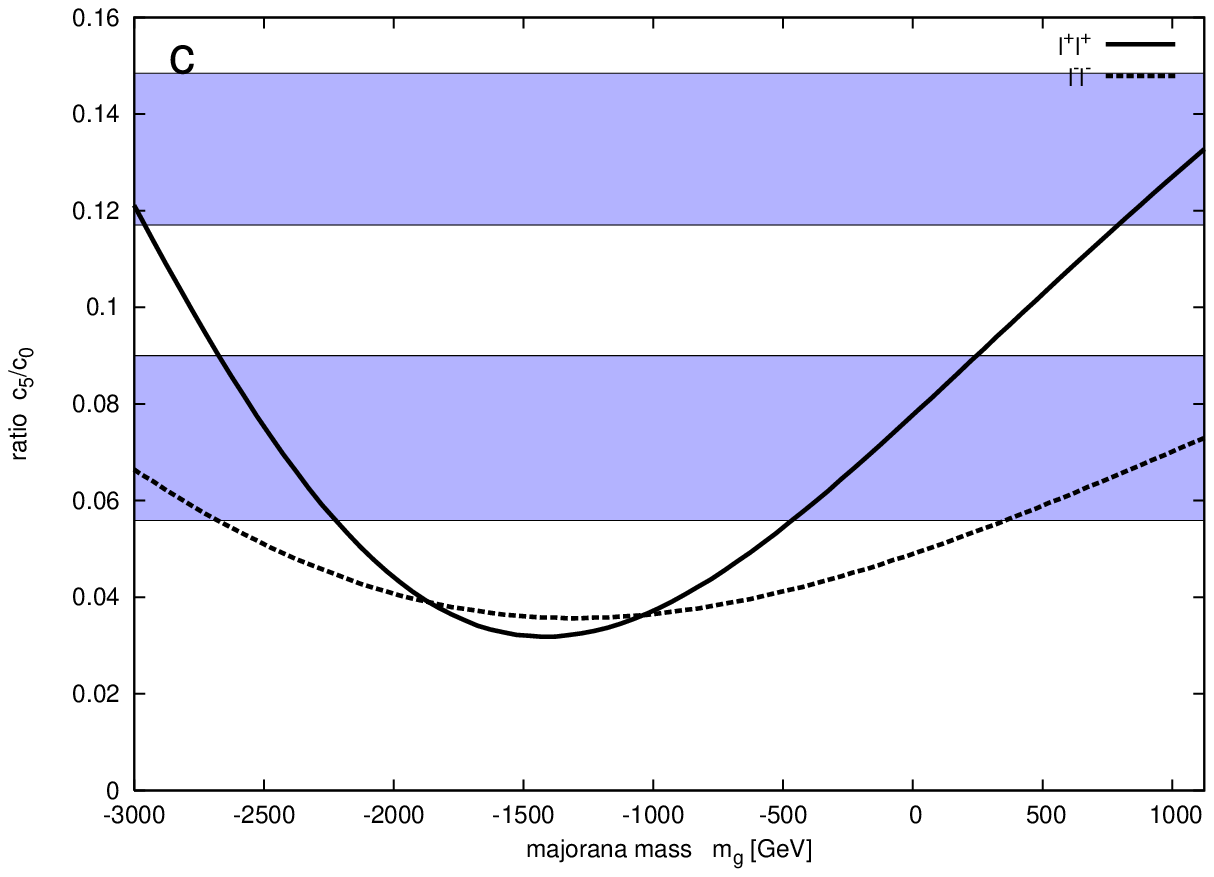}
\end{minipage}
\hfill
\begin{minipage}{8.0cm}
\includegraphics[scale=0.6]{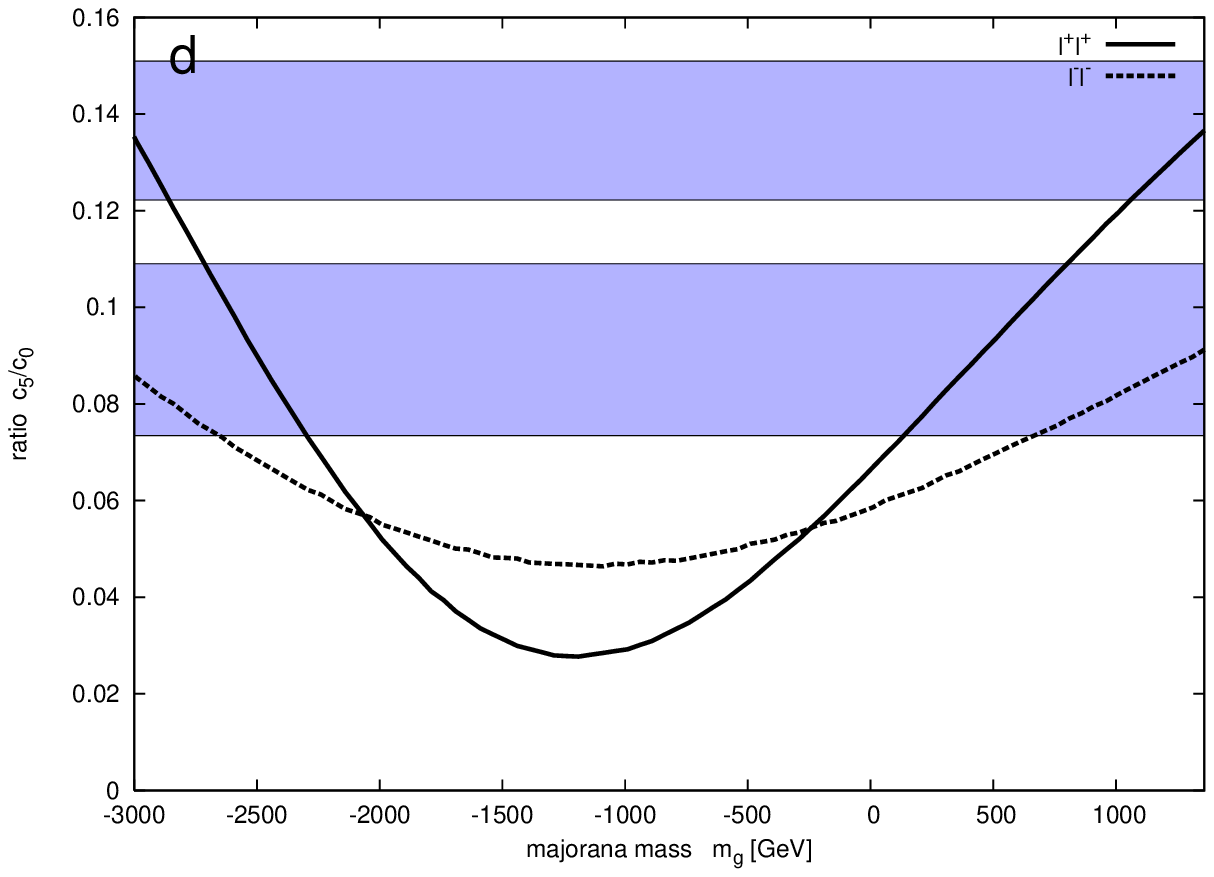}
\end{minipage}
\caption{$N(c_5)/N(c_0)$ dependence on the majorana gluino mass $m_g$ 
at a) Point A,
b) Point B,
c) Point C and
d) Point D.
Gray zones show 1$\sigma$ statistical errors
 for 3$\times 10^5$ events.}
\label{ratiohikaku}
\end{figure}

As discussed in the previous section,
the uncertainties on the leptonic branching ratios,
PDF, QCD NNLO corrections and squark mass errors
partly cancel by taking the ratio $N(l^+ l^+ ;c_5)/N(l^+l^+;c_0)$.
Roughly speaking,
$N(l^\pm l^\pm ;c_5)/N(l^\pm l^\pm;c_0)$ depends linearly on
$\sigma(\sq_L^+\sq_L^+)$/$\sigma(\gluino\sq_L)$.
In this model,
the ratio $\sigma(\sq_L^\pm\sq_L^\pm)/\sigma(\gluino\sq_L^\pm)$ decreases
as the majorana gluino mass decreases from the MSSM value.
The ratios $N(l^\pm  l^\pm  ;c_5)/N(l^\pm l^\pm ;c_0)$ 
as a function of the majorana gluino mass
are plotted for Points A$\sim$D in Figure \ref{ratiohikaku}a$\sim$d.
We show only the statistical error of
$\pm 1\sigma$ in the MSSM limit for $3\times 10^5$ events with a gray zone.
For example,
the ratio $N(l^\pm  l^\pm  ;c_5)/N(l^\pm l^\pm ;c_0)$ 
has a statistically significant difference 
from Pont B for $m_g \le 850$\,GeV at the 1$\sigma$ level. Because $\sigma(\sq^-\sq^-) \ll \sigma(\sq^+\sq^+)$ 
the sensitivity to $l^-l^-$ is worse.


\subsection{The Littlest Higgs model with T-parity}

We now consider the case where the $E\slush_T$ signature 
arises from decays of the quark partner $q_-$ in the LHT model.
The $q_-$ has a similar decay pattern to the $\sq$.
Indeed, the leptonic branching ratio of $q_-^+$ is $\sim$15\%,
which is almost the same as that of $\sq^+$ at Point C and D.
The acceptance should be similar to that of the MSSM,
because the acceptance depends on the decay kinematics,
namely on the mass difference between $\sq$ or $q_-$ mass and
the lightest R-odd or T-odd particle.

While the collider signal is similar to that of $\sq\sq$ production in the MSSM, 
the LHT model predicts different production cross sections 
from that of the MSSM.
The cross section 
$\sigma(q^+_- q^+_-)$ is $0.70\,$pb
and $\sigma(q^-_- q^-_-)$ is $0.15\,$pb
for $M_{q_-} \sim 800\,$GeV.
On the other hand,
$\sigma(\sq^+_L\sq^+_L)$ is $0.13\,$pb 
and $\sigma(\sq^-_L\sq^-_L)$ is $0.037\,$pb at Point C 
($m_{\sq} \sim 800\,$GeV).
For $M_{q_-}=1000\,$GeV,
$\sigma(q^+_- q^+_-)$ is $0.22\,$pb
and  $\sigma(q^-_- q^-_-)$ is $0.045\,$pb,
while $\sigma(\sq^+_L\sq^+_L)$ is $0.049\,$pb 
and $\sigma(\sq^-_L\sq^-_L)$ is $0.012\,$pb
at Point D ($m_{\sq}\sim 1000\,$GeV).
$\sigma(q^\pm_- q^\pm_-)$ is about 4$\sim$5 times larger than 
$\sigma(\sq^\pm_L\sq^\pm_L)$ at each point.
Note that $\sigma(\sq_L \sq_L)$ is always significantly smaller than 
$\sigma(q_- q_-)$ of the LHT model
no matter how heavy or light the $\gluino$ is.
Moreover,
there is no $\gluino$ production in the LHT model.
If the excess of the production cross section is established 
and the existence of a light gluino is excluded,
we can claim the LHT signature is observed.

At Point C,
$N(l^+l^+;c_0) = 467$,
$N(l^-l^-;c_0) = 233$,
$N(l^+l^+;c_5) = 62$ and 
$N(l^-l^-;c_5) = 17$
for an integrated luminosity of 86\,fb$^{-1}$.
To study $N(l^+l^+)$ or $N(l^-l^-)$,
we need to simulate all LHT production processes,
which is beyond the scope of this paper.
We assume the number of signal events from $q_-q_-$ production process 
is the number of signal events from $\sq\sq$ production process
scaled by the ratio of cross sections 
$\sigma(q_- q_-)/\sigma(\sq_L\sq_L)$,
and the other production processes are ignored for simplicity.
We expect 
$N(l^+l^+;c_0) \sim 816$,
$N(l^-l^-;c_0) \sim 162$,
$N(l^+l^+;c_5) \sim 302$ 
and $N(l^-l^-;c_5) \sim 49$ 
in the LHT model of $M_{q_-} \sim 800\,$GeV
for the same integrated luminosity.

At Point D,
we find $N(l^+l^+;c_0) =$ 571,
$N(l^-l^-;c_0) =$ 263,
$N(l^+l^+;c_5) =$ 78
and $N(l^-l^-;c_5)$ $=$ 24
for an integrated luminosity of 307fb$^{-1}$.
We expect
$N(l^+l^+;c_0) \sim 962$,
$N(l^-l^-;c_0) \sim 210$,
$N(l^+l^+;c_5) \sim 321$ 
and $N(l^-l^-;c_5) \sim 61$
in the LHT model of $M_{q_-} \sim 1000\,$GeV
for the same integrated luminosity.
We can see that in the LHT models,
$N(l^+l^+;c_0):N(l^-l^-;c_0)$ would be about 4:1.
This is different from the cases at the MSSM models Point C and D.

If there is no particle production except $q_-$
the ratio $N(l^+ l^+ ;c_5)/N(l^+l^+;c_0)$ is expected to be around 0.37 
in the LHT model of $M_{q_-} \sim 800\,$GeV,
which is larger than the value at Point C (0.133).
The ratio $N(l^+ l^+ ;c_5)/N(l^+l^+;c_0)$ is expected to be around 0.33
for $M_{q} = 1000\,$GeV,
and the value is larger than that at Point D (0.137).

Finally we comment on the case where $\sq$ and $\gluino$ are highly degenerate
so that we cannot detect the jets from $\gluino \to \sq q$ decay 
by hemisphere analysis.
Note that $\sigma(\gluino\sq) \gg \sigma(\sq\sq)$,
therefore the rate of the SS2$l$ events
could be as large as the LHT prediction.
However, even in this case,
$N(l^+l^+ ;c_0):N(l^-l^- ;c_0)$ in the MSSM is $\sim$ 2:1 and cannot be 4:1 
because there are $\gluino\sq$ and $\gluino\gluino$ contributions.

\section{Conclusion}

Information on the fundamental Lagrangian of SUSY model
can be extracted from each SUSY production process at LHC.
For example,
$\sq_L\sq_L$ production cannot occur 
without a majorana gluino mass,
because a chirality flip is required.
Thus the majorana nature of gluino mass can be extracted from this process.

This process can be investigated using the SS2$l$ events 
because ${\cal BR}(\sq_L^\pm \to l^\pm + X) \gg {\cal BR}(\sq_L^\pm \to l^\mp + X)$.
At LHC, however,
mixed production with other sparticles $\gluino$, $\sq$, etc. make it difficult 
to interpret the signal.
In particular, $\gluino \sq$ production 
also contributes to the SS2$l$ channel.

In this paper,
we have discussed a systematic method to separate
the production modes.
When we measure $\sigma(\sq_L\sq_L \to l^\pm l^\pm + X)$ in the MSSM,
we suffer from
a problematic background from $\gluino\sq_L$ production.
We have proposed a new method based on a hemisphere analysis 
as a solution to this problem.
In the hemisphere analysis, 
we assign high $p_T$ objects into two hemispheres, 
where each hemisphere contains high $p_T$ objects from the same parent particle 
with high probability.
Then we require 
that
there is only one jet with $p_T > 50\,$GeV 
in a hemisphere.
For the sample after some basic cuts,
30$\sim$40\% for $\sq_L\sq_L$ and
1$\sim$3\% for $\gluino \sq_L$ remain
after the hemisphere cut.
Therefore we can obtain SS2$l$ events
with enhanced $\sq_L\sq_L$ contribution,
which may be used to estimate $\sigma(\sq_L\sq_L)$.

We have also discussed two models
which have similar collider signals to the MSSM
but whose relevant production cross sections
are dramatically different.

One of the models is the MSSM with an extended gluino sector,
where the gluino can have a Dirac mass with an adjoint fermion $\ti a$.
$\sigma(\sq_L\sq_L)$ is 
sensitive to the fraction of majorana mass terms
in the gluino mass.
In the case where the gluino is pure Dirac,
$\sq_L\sq_L$ production cross section becomes zero.

We have applied our analysis 
to the model points with an extended gluino sector which have
the same mass spectra of some MSSM model points 
except for an additional heavy adjoint particle.
We estimate the number of SS2$l$ events as a function of the majorana gluino mass,
and estimate the sensitivity to the Dirac gluino mass.
We take only the statistical error into account
and assume that the
masses of the squarks and the branching ratios are known.
We find that the ratio $\sigma(\sq^+_L\sq^+_L)/\sigma(\gluino\sq_L^+)$
that can be estimated from the acceptance 
under the hemisphere cut is useful
because this quantity should be less sensitive to the error on 
the parameters and the acceptance and 
uncertainties of the PDF and QCD corrections.

We have also considered the LHT model.
In this model,
a set of T-odd partners is introduced to the SM matter particles and the EW gauge bosons.
The quark partner production cross section $\sigma(q_-q_-)$ 
is 4$\sim$5 times as large as 
the production cross section $\sigma(\sq_L\sq_L)$ in the MSSM.
The $q_-$ and $\sq$ have similar decay patterns and branching ratios.
Thus,
the number of SS2$l$ events from $q_-q_-$ productions in the LHT model
is expected 4$\sim$5 times larger
if patterns of the mass spectra are the same.
Moreover,
the LHT model has no process corresponding to
the process $\gluino\sq$ nor $\gluino\gluino$ production in the MSSM.
To exclude gluino productions is important to identify the LHT model.
This can be done by investigating $N(l^+l^+)/N(l^-l^-)$.

It is generally important to 
measure the production cross sections of sparticles separately
to verify the MSSM and distinguish various models.
In this paper,
we develop a method to identify gluino and squark production separately for SS2$l$ channel.
The method is based on the cuts on the kinematical configulation of the jets
and can be applied to the other models.
More development is needed for the model independent study of physics beyond the SM.

\section*{Acknowledgement}
This work is supported in part by the Grant-in-Aid for Science Research, 
Ministry of Education, Culture, Sports, Science and Technology, 
Japan (No.16081207, 18340060 for M.M.N.).

\newpage
\appendix
\section{Appendix}
\subsection{Mass spectra at our model points}
\label{appen1}

We show mass spectra for the selected model points
which have been analyzed.
They are calculated by ISAJET\,7.72. 
They are all mSUGRA mass spectra except that
gluino masses of Points C and D are 300\,GeV larger than mSUGRA predictions.

\begin{table}[htbp]
\center{
\begin{tabular}{|c||rrrrr|}
\hline
 mass parameter &Point A & Point B  & Point C  & Point D  &  SPS1a \cr
\hline
\hline
$m_0$&100&100&370&400&100 \cr
$m_{\h}$&340&450&340& 450&250 \cr
$A_0$&0&0&0&0& $-100$\cr
$\tan\beta$&10&10&10&10& 10\cr
sign\,$\mu$&+&+&+&+&+\cr
\hline
\multicolumn{6}{c}{}\cr
\hline
particle     &Point A & Point B  & Point C  & Point D  &  SPS1a \cr
\hline
\hline
$\gluino    $&    809.86&   1047.83&   1123.23&   1360.22&    595.19\cr
\hline
$\ti d_L       $&    741.58&    954.53&    812.70&   1021.11&    543.04\cr
$\ti u_L       $&    737.25&    951.16&    808.67&   1017.91&    537.25\cr
$\ti d_R       $&    712.94&    916.48&    786.98&    986.09&    520.14\cr
$\ti u_R       $&    714.56&    919.51&    787.79&    988.10&    520.45\cr
$\ti b_1       $&    683.97&    883.16&    731.10&    928.37&    491.92\cr
$\ti t_1       $&    559.18&    734.57&    585.39&    804.20&    379.14\cr
$\ti b_2       $&    708.33&    909.55&    776.58&    973.96&    524.58\cr
$\ti t_2       $&    738.66&    929.69&    780.92&    946.08&    574.64\cr
\hline
$\ti e_L       $&    256.36&    324.47&    437.13&    503.38&    202.12\cr
$\ti e_R       $&    168.27&    201.70&    393.09&    435.59&    143.00\cr
$\ti {\nu}_e      $&    243.67&    314.44&    429.70&    496.84&    186.00\cr
$\ti {\tau}_1     $&    160.79&    194.28&    387.18&    429.40&    133.39\cr
$\ti {\tau}_2    $&    258.50&    325.66&    437.28&    503.05&    206.02\cr
$\ti {\nu}_{\tau}    $&    242.89&    313.50&    428.00&    495.00&    185.06\cr
\hline
$\none      $&    132.74&    179.11&    133.95&    180.54&     96.05\cr
$\ntwo      $&    253.87&    345.39&    256.39&    348.15&    176.80\cr
$\nthre    $&   $-448.66$& $  -575.60$& $  -451.46$& $  -577.05$& $  -358.82$\cr
$\nfour     $&    467.28&    591.37&    470.52&    592.66&    377.84\cr
$\ch_1^+    $&    254.13&    345.95&    256.71&    348.72&    176.37\cr
$\ch_2^+    $&    466.48&    590.74&    469.74&    592.56&    378.26\cr
\hline
\end{tabular}
\caption{Mass spectra of sparticles for the selected model points 
}}
\end{table}
\newpage
\subsection{Branching ratios at our model points}
\label{appbranch}

We also show the branching ratios of the selected points.
They are also calculated by ISAJET.
Squarks mainly decay into charginos $\ch^\pm$ and the second lightest neutralino $\ntwo$ 
at each point.
Note that, at Points C and D,
$\ntwo$ does not decay into sleptons.
On the other hand, at Points A, B and SPS1a,
$\ntwo$ decays into $\none l^+l^-$ or $\none \tau^+\tau^-$.

\begin{table}[htbp]
\begin{tabular}{|cl||r|r|r|r|r||cl||r|r|r|r|r|}
\hline
 \multicolumn{2}{|c||}{mode} & \multicolumn{5}{c||}{BR(\%)} &
 \multicolumn{2}{c||}{mode} & \multicolumn{5}{c|}{BR(\%)}  \\
\cline{3-7}\cline{10-14}
 &  & \multicolumn{1}{c|}{    A} & \multicolumn{1}{c|}{   B} & \multicolumn{1}{c|}{   C} & \multicolumn{1}{c|}{   D}&\multicolumn{1}{c||}{SPS1a}
 & & & \multicolumn{1}{c|}{    A} & \multicolumn{1}{c|}{   B} & \multicolumn{1}{c|}{   C} & \multicolumn{1}{c|}{   D}&\multicolumn{1}{c|}{SPS1a} \\
\hline
\hline
$ \gluino    $ & $ \to \sq_L^+ q$ & $11$ & $10$ & $14$ & $13$ &12&$\ti t_1$ & $ \to \cpl_1 b$ & $49$ & $39$ & $43$ & $6.9$ &73\\
$            $ & $ \to \sq_L^- q$ & $11$ & $10$ & $14$ & $13$ &12&$$ & $  \to \cpl_2 b$ & $13$ & $21$ & $20$ & $41$ &0\\
$            $ & $ \to \sq_R   q$ & $38$ & $36$ & $32$ & $32$ &41&$$ & $  \to \none t$ & $22$ & $26$ & $22$ & $24$ &18\\
$            $ & $ \to \ti t_1 \bar{t}(\ti t_1^\ast t) $ & $7.1$ & $9.6$ & $5.3$ & $7$ &4.1& $$ & $  \to \ntwo t$ & $15$ & $14$ & $14$ & $3$ &9.5\\
$            $ & $ \to \ti b_1 \bar{b}(\ti b_1^\ast b) $ & $7.7$ & $6.9$ & $5.1$ & $5$ &8.9& $$ & $  \to \nthre t$ & $0$ & $0$ & $0$ & $15$ &0\\
$            $ & $ \to \ti b_2 \bar{b}(\ti b_2^\ast b) $ & $5.5$ & $5.2$ & $4.2$ & $4.2$ &4.9& $$ & $  \to \nfour t$ & $0$ & $0$ & $0$ & $10$ &0\\
\cline{8-14}
$$ & $ \to \ti t_2 \bar{t}(\ti t_2^\ast t) $ & $0$ & $0$ & $5.4$ & $4.5$ &0& $\ti t_2$ & $ \to \cpl_1 b$ & $23$ & $25$ & $22$ & $41$ &19\\
\cline{1-7}
$ \ti u_L $ & $ \to \cpl_1 d   $ & $65$ & $65$ & $64$ & $64$ &65& $$ & $  \to \cpl_2 b$ & $16$ & $12$ & $1.5$ & $2.8$ &22\\
$         $ & $  \to \cpl_2 d  $ & $1.6$ & $1.4$ & $1.9$ & $1.5$ &1.2& $$ & $  \to \ntwo t$ & $9.7$ & $11$ & $9.5$ & $18$ &7.6\\
$$ & $  \to \none u   $ & $0$ & $1.1$ & $0$ & $1$ &0.6& $$ & $  \to \nthre t$ & $8.5$ & $10$ & $11$ & $18$ &3.7\\
$         $ & $  \to \ntwo u   $ & $32$ & $32$ & $32$ & $32$ &32& $$ & $  \to \nfour t$ & $24$ & $23$ & $25$ & $19$ &18\\
$  $ & $  \to \nfour u   $ & $1.2$ & $0.9$ & $1.4$ & $1.1$ &1.0& $$ & $  \to \ti t_1 X$ & $17$ & $17$ & $17$ & $0$ &27\\
\hline
$ \ti d_L $ & $ \to \cm_1 u   $ & $61$ & $62$ & $60$ & $62$ &61& $\cm_2$ & $  \to \cm_1 X$ & $44$ & $48$ & $54$ & $55$ &41\\
$         $ & $  \to \cm_2 u    $ & $4.4$ & $3.3$ & $5$ & $3.6$ &4.1& $$ & $  \to \nt W^-$ & $37$ & $38$ & $45$ & $42$ &36\\
$$ & $  \to \none d   $ & $2$ & $1.8$ & $1.9$ & $1.7$ &2.4& $$ & $ \to l^- X$ & $12$ & $6.7$ & $0$ & $0$ &14\\
$         $ & $  \to \ntwo d   $ & $31$ & $32$ & $31$ & $31$ &31& $$ & $\to \tau^- X$ & $7.2$ & $5.4$ & $0$ & $0$ &6.6\\
\cline{8-14}
$$ & $\to \nfour d $ & $1.6$ & $1.2$ & $1.9$ & $1.4$ &1.4& $\cm_1$ & $  \to \none W^-$ & $21$ & $7.7$ & $100$ & $100$ &1.1\\
\cline{1-7}
$\ti b_1$ & $ \to \cm_1 t$ & $38$ & $36$ & $33$ & $38$ &43& $$ & $  \to \none l^-\bar{\nu}_l$ & $24$ & $54$ & $0$ & $0$ &0.4\\
$$ & $  \to \cm_2 t$ & $24$ & $28$ & $31$ & $37$ &0& $$ & $  \to \none \tau^-\bar{\nu}_\tau$ & $54$ & $38$ & $0$ & $0$ &98\\
\cline{8-14}
$$ & $  \to W^-  \ti  t_1$ & $10$ & $12$ & $12$ & $0$ &14& $\ntwo$ & $\to \none X$ & $45$ & $55$ & $100$ & $100$ &0.8\\
$$ & $  \to \ntwo b $ & $24$ & $21$ & $20$ & $22$ &36&$$ & $\to \none l^+l^-$ & $7.6$ & $18$ & $0$ & $0$ &13\\
\cline{1-7}
$\ti b_2$ & $ \to \cm_1 t$ & $15$ & $14$ & $4.9$ & $4.4$ &21& $$ & $\to \none \tau^+\tau^-$ & $46$ & $20$ & $0$ & $0$ &87\\
\cline{8-14}
$$ & $  \to \cm_2 t$ & $34$ & $35$ & $39$ & $41$ &0& $\nthre$ & $\to \none X$ & $14$ & $13$ & $14$ & $14$ &13\\
$$ & $  \to W^-  \ti  t_1$ & $13$ & $13$ & $13$ & $0$ &35& $$ & $\to \ntwo X$ & $24$ & $26$ & $25$ & $27$ &23\\
$$ & $  \to \none b$ & $19$ & $21$ & $23$ & $31$ &15& $$ & $\to \ch_1^\pm W^\mp$ & $29$ & $29$ & $30$ & $30$ &60\\
\cline{8-14}
$$ & $  \to \ntwo b $ & $9.6$ & $8.3$ & $3$ & $2.5$ &17& $\nfour$ & $\to \none X$ & $21$ & $19$ & $12$ & $13$ &8.5\\
$$ & $  \to \nthre b$ & $3.7$ & $3.5$ & $7.9$ & $9.4$ &5.4& $$ & $\to \ntwo X$ & $18$ & $22$ & $22$ & $25$ &15\\
$$ & $  \to \nfour b$ & $5.6$ & $5.2$ & $9.6$ & $11$ &7.4& $$ & $\to \ch_1^\pm W^\mp$ & $26$ & $26$ & $32$ & $30$ &52\\
\hline
\end{tabular}
\caption{Branching ratios of sparticles for our model points: 
here, $X$ means some SM particles.}
\end{table}

\clearpage
\subsection{Dominant processes contributing to SS2$l$ events}
\label{dominantcontribution}

Here we show the numbers of SS2$l$ events from each production process at Point B.
Note that SS2$l$ ($l^\pm l^\pm$) signals come not only from $\sq^\pm_L \sq^\pm_L$ production
but also from $\sq^+_L \sq^-_L$ (mainly $u_Ld_L$) production.
For $l^+l^+$ events, about 6\% of total $l^+l^+$ events are from $\sq\sq$ production.
The $l^+l^+$ events from $\sq_L\sq_L^\ast$ production are 10\% of
total $l^+l^+$ events from $\sq\sq$ production.

\begin{table}[htbp]
\center{
\small
\begin{tabular}{|c|c||r|rrrrrrr|}
\hline 
\multicolumn{2}{|c||}{Point B} & all & $c_0$ & $c_1$  & $c_2$                                                           & $c_3$ & $c_4$ & $c_5$ & ratio \cr
\hline 
\multicolumn{2}{|l||}{$l^+ l^+ $\ \ \ \ \ $\sq\sq$}&  1479&  1014&  1001&   894&   363&   883&   361& 0.356\cr
\cline{2-10}
\ \ \ \ \ \ &      $\ti u_L$$\ti u_L$&  1161&   788&   778&   701&   291&   692&   289& 0.367\cr
&      $\ti u_L$$\ti c_L$&    95&    68&    68&    58&    22&    58&    22& 0.324\cr
&      $\ti u_L$$\ti d_L$&    84&    58&    56&    51&    17&    50&    17& 0.293\cr
& $\ti u_L$$\ti d_L^\ast$&    90&    62&    61&    54&    19&    53&    19& 0.306\cr
& $\ti u_L$$\ti s_L^\ast$&    48&    37&    37&    29&    13&    29&    13& 0.351\cr
\hline
\multicolumn{2}{|l||}{$l^+ l^+ $\ \ \ \ \ $\gluino\sq$}&  1765&  1098&   433&   613&    52&   279&    36& 0.033\cr
\cline{2-10}
&      $\gluino$$\ti u_L$&  1581&   974&   388&   538&    41&   252&    30& 0.031\cr
&      $\gluino$$\ti c_L$&    22&    18&     8&     8&     2&     3&     1& 0.056\cr
&      $\gluino$$\ti d_L$&    32&    22&     8&    11&     2&     5&     1& 0.045\cr
& $\gluino$$\ti d_L^\ast$&    78&    53&    17&    38&     6&    12&     3& 0.057\cr
& $\gluino$$\ti s_L^\ast$&    39&    22&     9&    13&     0&     6&     0& 0\cr
\hline
\multicolumn{10}{c}{}\cr
\hline 
\multicolumn{2}{|c||}{Point B} & all & $c_0$ & $c_1$  & $c_2$                                                           & $c_3$ & $c_4$ & $c_5$ & ratio \cr
\hline 
\multicolumn{2}{|l||}{$l^- l^- $\ \ \ \ \ $\sq\sq$}&        519&   333&   323&   293&   123&   285&   122& 0.366\cr
\cline{2-10}
&   $\ti d_L$$\ti d_L$&        280&   188&   181&   164&    74&   159&    73& 0.388\cr
&   $\ti d_L$$\ti s_L$&         87&    62&    61&    56&    31&    55&    31& 0.5\cr
&   $\ti u_L$$\ti d_L$&        106&    67&    65&    57&    16&    55&    16& 0.239\cr
&   $\ti d_L$$\ti u_L^\ast$&    20&     7&     7&     7&     0&     7&     0& 0\cr
&   $\ti d_L$$\ti c_L^\ast$&    10&     7&     7&     7&     1&     7&     1& 0.143\cr
\hline
\multicolumn{2}{|l||}{$l^- l^- $\ \ \ \ \ $\gluino\sq$}&   881&   553&   241&   310&    23&   172&    15& 0.027\cr
\cline{2-10}
&   $\gluino$   $\ti d_L$&      615&   396&   184&   232&    21&   132&    14& 0.035\cr
&   $\gluino$   $\ti s_L$&       30&    18&     8&     9&     0&     5&     0& 0\cr
&   $\gluino$   $\ti u_L$&      135&    75&    23&    40&     2&    17&     1& 0.013\cr
&   $\gluino$   $\ti u_L^\ast$&  67&    43&    19&    21&     0&    13&     0& 0\cr
&   $\gluino$   $\ti c_L^\ast$&  20&    12&     4&     6&     0&     3&     0& 0\cr
\hline
\end{tabular}
\caption{Contributions from each production process to SS2$l$ events at Point B.}
}
\end{table}

\end{document}